\documentclass{aa}  

\usepackage{graphicx}
\usepackage{txfonts}
\usepackage{placeins}
\usepackage{lipsum}

\usepackage[T1]{fontenc}
\usepackage{ae,aecompl}
\usepackage{enumitem}
\usepackage{float} 
\usepackage[usenames,dvipsnames,table,xcdraw]{xcolor}
\usepackage{xcolor}
\usepackage{xspace}
\usepackage[para]{threeparttable} 
\usepackage{multirow}
\usepackage{multicol}
\usepackage{pdfpages}
\usepackage[version=4]{mhchem}

\usepackage{afterpage}

\usepackage{subcaption}
\usepackage[colorlinks=true,linkcolor=Blue,citecolor=Blue,urlcolor=black,anchorcolor=black,bookmarks=true,breaklinks=true,hypertexnames=false,bookmarksnumbered=true]{hyperref}
\begin{document} 

\newcommand{\vdag}{(v)^\dagger}
\newcommand\aastex{AAS\TeX}
\newcommand\latex{La\TeX}

\newcommand{\todo}[1]{\textcolor{red}{todo:#1}}
\newcommand{\solar}{$_\odot$}
\newcommand{\zml}{$Z = 10^{-4}$}
\newcommand{\zmu}{$Z = 0.03$}
\newcommand{\solarperyr}{$_\odot \, \textrm{yr}^{-1}$}
\newcommand{\bse}{\texttt{bse}}
\newcommand{\binaryc}{\texttt{binary\_c}}
\newcommand{\nupycee}{\texttt{NuPyCEE}}

\newcommand{\Mdot}{$\dot{M}$}
\newcommand*\diff{\mathop{}\!\mathrm{d}}
\newcommand{\Z}{$Z$}
\newcommand{\tento}[1]{$10^{#1}$}
\newcommand{\timestento}[2]{$#1 \times 10^{#2}$}
\newcommand{\iso}[2]{\hbox{${}^{#1}{\rm #2}$}}

\newcommand{\angbrack}[1]{$\langle${#1}$\rangle$}
\newcommand{\sbs}[2]{$#1_{\rm #2}$}
\newcommand{\Yeff}{$Y_{\rm eff}$}
\newcommand{\Yprim}{$Y_{\rm prim}$}
\newcommand{\Ysecavg}{$\langle Y_{\rm sec}\rangle$}
\newcommand{\Ysec}{$Y_{\rm sec}$}

\newcommand{\Ysing}{$Y_{\rm single}$}

\newcommand{\q}{$q$}

\title{Binary stellar evolution yields in galactic chemical evolution calculations}

   \subtitle{}

   \author{
   Alex Kemp\inst{1}\thanks{
    \email{alex.kemp@kuleuven.be}}
    \and
    Tejpreet Kaur\inst{2}
  }

   \institute{Institute for Astronomy (IvS), KU Leuven, Celestijnenlaan 200D, 3001, Leuven, Belgium
   \and
   Department of Space, Planetary \& Astronomical Sciences and Engineering, Indian Institute of Technology Kanpur, Kanpur 208016, India}

   \date{}

 
  \abstract
   {Galactic chemical evolution (GCE) models aim to bring together stellar yields and galactic evolution models to make predictions for the chemical evolution of real stellar environments. Until recently, stellar yields accounting for binary stellar evolution were unavailable, leading to an inability for GCE calculations to account for most binary stellar evolution effects. Fortunately, effective stellar yields that account for binary stellar evolution at a population level can be pre-computed and then used as if they were single yields.}
   {We present a framework for the computation of effective stellar yields that accounts for a mixed population of binary and single stars under an adjustable mix of binary evolution settings: the binary fraction, the accretion efficiencies of winds, Roche-lobe overflow, and supernovae.  We emphasise the critical need for more complete yield coverage of the binary nucleosynthesis and evolution, without which the ability to make accurate predictions on the true role of binarity on GCE calculations is hamstrung. We also provide clear guidelines for future stellar modelling works to ensure their results are maximally useful to the wider community.}
   {We compute effective stellar yields using detailed binary stellar yields accounting for binary induced mass-loss from a solar-metallicity donor star. We study the effect of varying the binary mixture and accretion efficiencies, and consider a range of models for the treatment of accreted material on the secondary in detail.}
   {In the absence of detailed binary yields for the secondary, we put forth a model for the composition of accreted material that preserves the signature of the primary's nuclear processing within the post-mass-transfer secondary yields. This model includes special treatment for isotopes of the light elements Li, Be, and B and accreted radioisotopes. Among the binary parameters, we find that the binary fraction, which determines the ratio of binary and single star systems, has the most significant effect on the effective stellar yields, with widespread impact across most isotopes. In contrast, varying the accretion efficiencies produces comparatively minor changes. We also find that the binary fraction has a significant influence on the logarithmic elemental abundance ratios relative to H present in the effective yield;   this impact is the largest for the lower-mass primaries.}
   {Comprehensive coverage of binary systems is essential for advancing our understanding of the role of binary stellar evolution in galactic chemical evolution. Priority areas include low-mass stellar evolution, binary mergers, and supernova yields coupled with the evolution of their binary progenitors with nuclear post-processing. The low-metallicity regime is also largely unexplored, offering great opportunity for novel and impactful research in this area.}


\maketitle
%

\section{Introduction}

The question of the origin of the elements touches on many different fields of physics, such as stellar and galactic evolution, nuclear physics, and cosmology. The goal is to predict how, when, and where the elements are produced, and in what quantities. These predictions are often informed by galactic chemical evolution (GCE) codes,\footnote{This is something of a misnomer, as the modelling is often applied to non-galactic stellar populations such as globular clusters or galactic sub-regions.} which model the chemical evolution of stellar populations using stellar yields computed by detailed stellar evolution models. Comparing these models with observations of stellar abundances (e.g. \citealt{wilson2019short,buder2021}) can expose gaps in our understanding of the underlying physics.

While there can be significant differences in the capabilities of different GCE codes in their treatment of galactic physics  (see e.g. \citealt{kobayashi2000,kobayashi2020origin,prantzos2018,prantzos2020,cote2017omega,prantzos2018, sahijpal2018, kaur2019}), 
they all rely on stellar yields. These yields are derived from detailed models of relevant stellar sources, and traditionally encompass yields from single low-mass stars (including AGB winds; see e.g. \citealt{karakas2010,fishlock2014,karakas2016,pignatari2016,cinquegrana2022}), single massive stars (including supernovae; see e.g. \citealt{woosley1995, kobayashi2006,pignatari2016,limongi2018}), and type Ia supernovae \citep{iwamoto1999,seitenzahl2013,seitenzahl2016,delosreyes2020manganese}. Additional sources of enriched material include cosmic ray spallation \citep{prantzos2012}, classical novae \citep{spitoni2018,grisoni2019,vasini2022,kemp2022li,bekki2024,kemp2024}, and binary neutron star mergers \citep{kobayashi2023nsm}.

There is a wide range of stellar yields for GCE modellers to choose from, each incorporating different stellar physics, such as internal stellar mixing, wind prescriptions, and rotation calculated over a range of metallicities. Modern GCE codes are well-equipped to use different sets of stellar yields to build populations of single stars, but their ability to account for binary stellar physics is much more limited \citep{cote2017omega, cote2018omegaplus, sahijpal2018, kaur2019}. Some can include binary phenomena such as type Ia supernovae, novae, and neutron star mergers by using delay time distributions from binary population synthesis simulations. However, most lack the ability to account for binary stellar evolution in normal stars or build a mixed population of binary and single star systems.

Previously, this was not a pressing issue because stellar yields accounting for binary stellar evolution were unavailable. However, this is no longer the case. 
\cite{farmer2023} present detailed massive stellar yields (including supernovae) for stars between 11 and 45 M\solar\ that have undergone Roche lobe overflow (RLOF, \citealt{paczynski1971binary}) onto a companion, while \cite{brinkman2023} present a sparser grid of stellar wind yields for stars between 10 and 80 M\solar\ undergoing similar interactions. Both of these works fix the initial mass ratio\footnote{The very recent work of \cite{nguyen2024} consider a range of mass ratios and orbital periods, but present wind yields for only a small number of elements.} of the binary, and primarily\footnote{\cite{brinkman2023} consider a range of initial orbital separations, some of which lead to `case A' (main sequence) mass transfer.} deal with case B mass transfer, where RLOF begins after the main sequence but before the onset of core He burning.

These currently available yield sets do not represent a complete model of binary stellar evolution. They account for the binary stellar evolution of the primary (initially more massive) donor star, but not the secondary (initially less massive) accretor; they are only available at solar metallicity. The yields of \cite{brinkman2023} models do not include supernovae, while the yields of \cite{farmer2023} are computed only for a single orbital configuration. Further, there are currently no complementary sets of detailed binary yields for stars below 10 M\solar\ or for more complicated binary stellar evolution channels, leaving a truly complete picture of binary stellar evolution obtainable only through reliance on rapid binary population synthesis codes with nucleosynthesis capabilities such as \binaryc\ \citep{izzard2004,izzard2006,izzard2018binary} and \texttt{BPASS} \citep{stanway2016,stanway2018}.

Nonetheless, these yield sets mark the opening of a new frontier in the exploration of binary stellar evolution, paving the way for more comprehensive future studies. It is, therefore, critical that our models for galactic chemical evolution are able to make effective use of these new data sets. Accounting for mixed populations of binary and single stars allows for more accurate and realistic GCE models. Further, this kind of modelling has the potential to provide valuable feedback to binary stellar physics in the same way it has done to single stellar models.

In this work, we lay out a framework for the calculation of effective stellar yields, a concept introduced in \citep{brinkman2019} to allow the inclusion of mixed populations of binary and single stars in existing GCE codes. These effective stellar yields can be treated by GCE codes as if they were simply single stellar yields, without internal modification. Further, they can be rapidly recalculated using different assumptions about binary birth distributions and mass transfer efficiencies for stellar winds, RLOF, and supernovae, allowing direct coupling between the GCE model and some of the assumptions about binary stellar physics. Using yields from \cite{farmer2023}, we demonstrate the effect that varying mass transfer efficiencies and binary fractions has on effective stellar yields. We also provide a checklist to guide future binary stellar modellers on the output needed to maximise the usefulness of their yield calculations and discuss where future modelling efforts should focus to form a more complete picture of how binary stellar evolution impacts stellar yield calculations.

In Section \ref{sec:effective_yields}, we describe the calculation of the effective stellar yields. In Section \ref{sec:res}, we describe and discuss the effective yields computed using binary yields from \cite{farmer2023}. In Section \ref{sec:checklist}, we offer a guide to future binary modelling works to improve functionality with GCE calculations and discuss the need for additional sets of binary stellar models. We conclude in Section \ref{sec:conc}.

\section{Effective stellar yields}
\label{sec:effective_yields}

The effective stellar yield is essentially a mixing equation for stellar populations, representing the average yield (considering different possibilities of secondary masses and, in principle, orbital periods) of a binary with a given initial primary mass. These yields inherently correct for the mass of star-forming material required to form the secondary, allowing them to be used like single stellar yields without any mass-conservation issues.

Eqn. 4 of \cite{brinkman2019} defines the effective stellar yield, \Yeff, as

\begin{equation}
    Y_{\rm eff} = \frac{(1-h)Y_{\rm single} + h(Y_{\rm prim} + \langle Y_{\rm sec}\rangle)}{1 + h\langle q \rangle},
    \label{eq:brink2019yeff}
\end{equation}

where \Yprim\ is the net yield\footnote{The difference between the mass of a given isotope initially present in the star and the mass of that isotope ejected into the ISM} of the primary, \angbrack{\Ysec} is the average net yield of the secondary with consideration to the secondary birth distribution, \Ysing\ is the net yield of a single star with equivalent mass and single stellar evolution to the primary, $h$ is the binary fraction relevant to the primary mass, and \angbrack{\q} is the average value of the mass ratio $q=M_{\rm sec}/M_{\rm prim}$ with consideration to the secondary birth distribution.

This equation naturally conserves the mass of star-forming material and, assuming the same initial mass function for single stars and primaries, can be initialised with the same frequency (or weighting) as a single star of the same mass. This allows a GCE code to treat a table of effective stellar yields as if it were a table of single yields.

\subsection{Calculation of terms for mass transfer onto a main sequence companion}

The effective stellar yield defined in equation \ref{eq:brink2019yeff} relies on a combination of binary parameters and stellar yields, the calculation of which determines the degree of realism of the calculation. Here, we discuss $h$,   \angbrack{\q}, \Yprim, and \angbrack{\Ysec} in detail, and how they relate to stellar physics.

\subsubsection{The binary fraction $h$ and the average mass ratio \angbrack{q}}

The birth distribution of secondary masses is commonly assumed to be flat in \q\ between 0 and 1, implying that \angbrack{\q}$=0.5$.\footnote{\cite{farmer2023} and \cite{brinkman2023} keep \q\ fixed (to 0.7 and 0.9, respectively) when initiating their binaries, but in nature a wide range of secondary masses are possible.} Usage of a more complicated distribution is trivial, although the distribution assumed for \q\ must also be propagated into the calculation of \angbrack{\Ysec}. In this work, we assume that the distribution of secondary masses is flat in \q, but it is worth considering the effect of the upper and lower bounds on this distribution.

The binary yields currently available are for a very simple evolution channel: stable mass transfer from an evolved donor onto a companion star. However, not all Roche-lobe filling giants undergo stable mass transfer. Many of these interacting stars will instead undergo unstable mass transfer leading to a common envelope resulting in either a merger with its companion or a significant shortening of the orbital period, often followed by further episodes of mass transfer.

Consider two different assumptions on the bounds for \q. Applying bounding values of 0 and 1 on \q, and assuming a binary fraction representative of nature, is equivalent to assuming that this simple evolution channel is representative of all binaries. This assumption is made in this work, and so propagates to our reported effective stellar yields. Alternatively, raising the lower limit of \q\ from 0 (while reducing the overall binary fraction accordingly) is equivalent to treating these un-modelled binary stellar evolution channels as single stars. Ideally, we would instead break down the `binary' term in the effective yield calculation further into multiple terms accounting for different binary evolution channels, with consideration to the birth distributions of the binary systems that undergo these channels. We discuss this in more detail later in Section \ref{sec:effective_yields_better}.

The binary fraction $h$ is widely accepted to depend on the birth mass of the primary, motivated by observational studies such as \cite{sana2012} and \cite{moe2017} which reveal lower binary fractions for low-mass stars than for those of high-mass stars such as those considered in \cite{farmer2023}. Any dependence on metallicity is uncertain and usually ignored, although it could be easily accounted for within this formalism since the metallicity of the underlying stellar models is always known. All a user would need to do would be to match detailed models computed at the relevant metallicity to the correct binary fraction.

The question of how close two stars need to be to be considered a binary is relevant but not often addressed. Ideally, the binary fraction used would be relevant to the binaries interacting as modelled by the underlying work that produced the yield sets. This could be accomplished in practice by introducing additional sub-divisions beyond simply `binary' and `single' (see Section \ref{sec:effective_yields_better}). Since we are currently restricted to a single binary evolution channel by the availability of binary stellar yield calculations, we cannot account for orbital period considerations in detail.

\subsubsection{\Ysing, \Yprim, and \Ysec}
 
The stellar yield terms \Yprim\ and \Ysec\ are where additional physics can start to be introduced in order to form a more realistic stellar population. \Ysing\ is simply the yield of a single star of equivalent mass, ideally using the same physical ingredients. A more realistic calculation of this term would include a mix of rotating and non-rotating stellar models. Examples of studies which have incorporated mixed populations of rotating and non-rotating stars include \cite{romano2019,prantzos2020}.

Consideration of \Yprim\ and \Ysec\ allows the introduction of mass transfer efficiencies, enabling this aspect of binary stellar physics to be probed even without grids of detailed stellar models varying these parameters. We shall address \Yprim\ first.

\cite{farmer2023} provide a breakdown of the primary yields in terms of the mass loss mechanism, rather than simply providing the total yield for the star. The absolute (total) yield of a given isotope for a given initial mass is

\begin{equation}
    Y_{\rm prim}= Y_{\rm winds} + Y_{\rm RLOF} + Y_{\rm SN},
\end{equation}

where \sbs{Y}{winds}, \sbs{Y}{RLOF}, and \sbs{Y}{SN} are the absolute yields from stellar winds, RLOF, and the supernova, respectively. The net yields can be calculated by subtracting the amount of each isotope initially present within the star from the absolute yield. Note that this is not identical to subtracting `init' columns in the yield tables of \cite{farmer2023}, which instead specify the mass of the isotope that would have been lost to the ISM had the composition of the lost material been identical to the birth composition. The subtraction of these `init' columns from the appropriate absolute yield columns, instead, gives the net enrichment (or depletion) of the material lost through each mass transfer mechanism. The sum of these terms does not consider that there is a non-zero remnant mass left behind at the end of the star's life, a `sink' term necessary to the computation of the true net yield.

The ability to easily compute the net enrichment or depletion through each mass transfer channel is extremely useful when building a statistical binary model accounting for the composition of the accreted material. As we discuss later, it opens the door both to precisely controlling mass transfer efficiency assumptions when calculating those binary yields. We strongly recommend future binary modelling works report their yields in a similar format to that of \cite{farmer2023}.

\cite{farmer2023} assume fully conservative mass transfer, meaning that all of the mass lost through RLOF is transferred to the companion. In contrast, \cite{brinkman2023} assume completely non-conservative mass transfer, meaning none of the mass lost from the primary goes to the companion, instead being lost to the ISM directly. This decision affects the exact orbital evolution of the binary during mass transfer, which can impact the mass-transfer shut-off point. For RLOF, we expect quite conservative mass transfer for close binaries \citep{demink2007,langer2020}, but it should be noted that the details of the orbital evolution will only be truly correct for the precise orbital configuration computed anyway. We also note that the results of \cite{brinkman2023} suggest that the precise orbital separation -- and by implication, the orbital evolution -- are of secondary importance compared to the evolution state of the donor and its initial mass.

When considering the primary yield, it is not necessary to enforce the mass transfer efficiency assumptions adopted in the underlying stellar models. Instead, the mass transfer efficiency for each type of mass transfer can be treated as a free parameter and the yield of the primary can be written as

\begin{equation}
    Y_{\rm prim}= (1-\beta_{\rm winds})\cdot Y_{\rm winds} + (1-\beta_{\rm RLOF})\cdot Y_{\rm RLOF} + (1-\beta_{\rm SN})\cdot Y_{\rm SN},
    \label{eq:beta_simple}
\end{equation}

where $\beta$ is the mass transfer efficiency (conservativeness). A value of one implies fully conservative mass transfer, where all material is transferred to the companion, while a value of zero implies no material is transferred, and all material from that mass transfer mechanism is lost directly to the interstellar medium (ISM). \sbs{\beta}{winds} will typically be low ($\lesssim 0.1$) for most binaries, excepting cases where a slow, dense wind is able to be funnelled to the secondary by the Roche potential (wind-RLOF, see \citealt{abate2013wind}). In contrast, values of \sbs{\beta}{RLOF} close to unity are expected for most stable mass transfer scenarios through RLOF, although RLOF occurring in simultaneously with very strong wind mass loss can in principle bring this number down. \sbs{\beta}{SN} should be essentially 0 in all cases; it is difficult to imagine a scenario where a stellar companion would accrete non-negligible amounts of material either during or shortly after a supernova explosion due to the extremely high ejecta velocities.

Using equation \ref{eq:beta_simple}, the yield from the primary can be modified to account for the desired accretion efficiency for each mass transfer mechanism. This is done without loss of consistency because, as far as the underlying stellar models are concerned, the material was removed, and the primary's subsequent evolution can be assumed to be indifferent to what happened to the material after said removal. We should note that this will not be generally true when considering a more complete model of binary stellar evolution, where multiple episodes of mass transfer (including `reverse' mass transfer from the secondary to the primary) can take place.

Finally, we come to the secondary yield. If we consider that the most common evolution phase for the accretor is a main sequence companion, then we are able to construct a well-motivated estimate for this yield. In a main sequence star, there is coupling between the core and envelope. This means that changes to the mass of the envelope leads to adjustments to the rest of the star -- including the core -- on a thermal timescale, which is short compared to the (nuclear) evolutionary timescale of the main sequence. After the material has been dumped onto and the accretor has adjusted, it is expected to evolve in a fashion similar to a star born with an initial mass equal to its mass after accretion, although tell-tale structural features can remain as evidence of its previous mass transfer \citep{henneco2024}.

Considering this, the yield for a secondary star born with initial mass \sbs{M}{2} that accreted mass $\Delta M$ on the main sequence would be equal to the yield of a star born with initial mass $M_{2}+\Delta M$. However, the accreted material may be enriched or depleted relative to the initial composition of the binary. Thus, we introduce the composition correction term \sbs{Y}{dump}, which can be written in terms of the reported yield of the primary and the assumed values of $\beta$ as

\begin{equation}
\begin{split}
    Y_{\rm dump}= &\beta_{\rm winds}\cdot (Y_{\rm winds}-Y_{\rm winds,\ init})\ +  \\
    &\beta_{\rm RLOF}\cdot (Y_{\rm RLOF}-Y_{\rm RLOF,\ init})\ + \\
    & \beta_{\rm SN}\cdot (Y_{\rm SN}-Y_{\rm SN,\ init}),
\end{split}
\end{equation}

where the  subscript `init' denotes the mass of the relevant isotope originally present in the lost material. This quantity is reported by \cite{farmer2023}, and is useful as it allows easy calculation of the net enrichment or depletion of material lost through each mass transfer channel.

\sbs{Y}{dump} can be thought of as the net production of the appropriate isotope that is dumped onto the secondary, preserving the nuclear processing that occurred in the primary before mass transfer. Without this quantity, we would have to assume accretion of birth-composition material, artificially reducing (or enhancing) the binary yields for many isotopes. Instead, we prefer the other extreme, which is that -- with some exceptions -- the net yields of the primary are preserved as modifications to the yield of the secondary. This means that, for example, if there is a net RLOF yield of 0.1 M\solar\ of C-12, and all of this mass is transferred to the secondary (i.e. $\beta_{\rm RLOF}=1$), then the C-12 yield of the secondary -- after correcting its yield for the raw amount of mass transferred -- will be increased by 0.1 M\solar. This assumption will be imperfect, and does not account for evolutionary differences due to the altered composition of the secondary. However, a more detailed treatment would require a wide range of post-accretion stellar models with yield calculations and this dataset, at present, does not exist.

Assuming the secondary has an identical birth composition to the primary, the material accreted onto the secondary is not altered after accretion, and evolutionary changes due to the altered composition are neglected, the yield for the secondary can be written as

\begin{equation}
    Y_{\rm sec}= Y_{\rm single}^{M_2+\Delta M} + Y_{\rm dump},
\end{equation}

where $Y_{\rm single}^{M_2+\Delta M}$ is the yield from a single star of mass equal to the post-transfer mass of the secondary.

Careful choice of the yield set used for \sbs{Y}{single} offers the chance to improve the physicality of the model. For example, yields sets of rotating stellar models such as \citep{limongi2018} or \citep{brinkman2021} could be used to study the effect of accounting for binary-induced spin-up. One might attempt to account for changes in the metallicity of the star by selecting a set of single yields with an altered birth composition reflective of the deposited material. However, changes to the bulk metallicity are likely to be small for the cases of stable mass transfer we are considering here, and the implicit assumption of full stellar mixing of the accreted material will also not be valid for most stars. Further, attempting to account for these secondary effects will likely necessitate using different yield sets from different studies, which may employ very different physics to the adopted binary yield set. These breaks in self-consistency may outweigh the benefits of attempting to model these kinds of second-order effects from the accretion on the secondary's evolution.

\sbs{Y}{dump} describes the net enrichment or depletion of a given isotope by the primary that was accreted by the secondary, and its use as a correction term ensures that this net yield is still released at the end of the secondaries life. However, this is patently false for radioisotopes and the fragile light elements Li, Be, and B.

Even assuming no nuclear processing of net-enriched accreted material on the secondary, radioisotopes will naturally decay during their time on the secondary, shifting the balance of the corrective \sbs{Y}{dump} term. The effects of this can, however, be quickly calculated provided that the time from accretion to the point of mass loss on the secondary is known. Which brings us to the question of stellar lifetimes, and how we can approximate their modification in the event of mass accretion.

We are interested in two time-intervals relative to the birth of the system: the lifetime of the secondary relative to the birth of the system, \sbs{t}{sec}, and the time between when the secondary accretes material, \sbs{t}{acc}, and when the secondary releases that material to the ISM. The age of the primary at the end of its life can be approximated as the time of accretion onto the secondary to within a few million years, due to the short post-main sequence lifetimes of massive stars. Similarly, the age of the secondary at the end of its life can be approximated as the time of the secondary's mass-loss. A more accurate timing could in principle be calculated using times of peak mass loss through winds and RLOF should this information be provided (see Section \ref{sec:checklist}), but this improved fidelity would have negligible impact on the binary's effective yields.

In order to estimate the lifetime of the main sequence secondary, we consider two effects of the mass accretion. The first is an increased rate of nuclear burning in the core, reducing the secondary's remaining lifetime, and the second is a structural adjustment that increases the core mass of the secondary. At the time of accretion, some amount of H has already been processed into He; it is this mass of processed material that we conserve during the age calculation. This mass can then compared with the final He core mass of a star with mass equal to the post-mass transfer mass to estimate the remaining stellar lifetime.

The fraction of the secondary's initial lifespan that has expired at the time of mass accretion is $t_{\rm acc}/t_{\rm single}$, where \sbs{t}{single} is the lifetime of a single star with the same initial mass of the secondary. This can then be used to estimate the mass fraction of H already processed at the time of accretion:

\begin{equation}
    M_{\rm H,proc@t_{\rm acc}} = M_{\rm c,single@M_2} \cdot t_{\rm acc}/t_{\rm single}.
\end{equation}

Here \sbs{M}{c,single} is the final H-depleted core mass of a single star with the same initial mass as the secondary. Note that this implicitly assumes a constant nuclear processing rate throughout the star's life, which is untrue. A more sophisticated approach would be to use detailed stellar evolution tracks to extract the actual amount of H processed at the time of mass accretion, but this would increase the computational effort -- and burden of reporting to binary stellar modellers -- significantly, and result in only minor improvement on a corrective term in the yield calculation.

Under the same assumption of constant nuclear processing, the mass of processed H, $M_{\rm H,proc@t_{\rm acc}}$, can then be used in conjunction with the final H-depleted core mass of a single star with mass equal to $M_2+\Delta M$, $M_{\rm c,single@M_2+\Delta M}$ to get the remaining lifetime of the secondary:

\begin{equation}
t_{\rm sec}-t_{\rm acc} = M_{\rm H,proc@t_{\rm acc}}/M_{\rm c,single@M_2+\Delta M}.
\end{equation}

This can then be used to calculate the secondary's lifetime, \sbs{t}{sec}. The remaining lifetime of the secondary after mass accretion, $t_{\rm sec}-t_{\rm acc}$, is the amount of time any accreted radioisotopes will have to decay before they and their decay products are released by the secondary. Hereafter we refer to this as the decay time. 

With knowledge of the amount of each radioisotope accreted onto the secondary and the decay time, it is now possible to calculate the state of \sbs{Y}{dump} at the time of the secondary's release accounting for radioactive decay. We accomplish this in practice by using the \texttt{radioactivedecay}\footnote{\href{https://pypi.org/project/radioactivedecay/}{https://pypi.org/project/radioactivedecay/}} \citep{malins2022} Python package, which can rapidly calculate the final quantities of each isotope accounting for complete radioactive decay chains. We use it in its default configuration, which includes decay information for 1252 radionuclides from 97 elements taken from  \cite{eckerman2008}.

Only the radioisotopes which have yields calculated by the underlying binary stellar models are relevant for initialising the calculation; further, we need only consider decay products that correspond to isotopes which have yields calculated by the underlying binary stellar models. Decay products that are not among the isotopes considered by the underlying stellar models are compounded into their relevant daughter isotope(s) or discarded. The quantities of these decay products are negligible compared to the amount of the daughter isotopes already in existence for all isotopes considered.

Both the primary and the secondary act to almost completely destroy the light elements Li, Be, and B. However, the initial abundances of these elements can only be destroyed once, and the stellar models used to compute the secondary yield assume a certain amount of these elements present. If we preserve the net depletion of these isotopes by the primary, the material transferred to the secondary will effectively be depleted twice, resulting in net-negative absolute yields for these elements from the secondary, which is unphysical. To prevent this, we reverse the depletion of these isotopes by the primary within the dumping term by transferring the net (negative) amount to the daughter particles of each isotope in the appropriate ratios (assuming they are destroyed in H-burning reactions, this is He-4 and a small amount of He-3). This resets the abundances of these isotopes within the dumping array while conserving mass. These restored accreted light isotopes are then almost completely destroyed by the secondary during its natural evolution, restoring the original outcome of light element destruction.

In this way, we can calculate the secondary yield for an individual binary accounting for evolution changes due to its mass gain, modifications to its yield stemming from the specifics of the material it accreted, and the radioactive decay of the radioisotopes present within this material. The calculation ultimately depends on both the primary mass, which determines the amount of mass transferred, and the secondary mass. Therefore the yield calculation must be done for a range of initial secondary masses for each primary mass, and then the results averaged with consideration to the previously discussed birth distribution of secondary masses to arrive at \angbrack{\sbs{Y}{sec}}.

Note that as the effective stellar yield is intended to replace single yields, it does not provide any information about any time offset between the primary and secondary yield releases (although that time is used to calculate more accurate yields by accounting for the radioactive decay of accreted material). Indeed, the time offset will be a function of the secondary mass, and so is poorly defined for the effective yield, which considers an average secondary yield. For massive stars, this time offset is typically only a few tens of millions of years and so will not be important for most science cases.

\subsection{Towards a more complete treatment of binary stellar evolution}
\label{sec:effective_yields_better}

The above is a detailed explanation of a method for calculating the effective stellar yield when considering the simplest possible binary stellar evolution channel: stable mass transfer from an evolved star to a main sequence companion. This channel is highly relevant in the context of the detailed binary stellar yields currently available, but it is to be hoped that stellar yields from more complex binary evolutionary channels will be available in the future. In this section, we address how these yields can be incorporated into the effective stellar yields formalism, and thereby included in galactic chemical evolution calculations.

The effective stellar yield is essentially a mixing equation for the yields from binary and single star systems, and as such it is extremely versatile. An arbitrary number of evolution channels can be incorporated by subdividing the binary (or single) population into different evolution channels. This can be achieved by expanding the binary (or single) term, breaking it up into different channels accompanied by an appropriate weighting factor.

Suppose that yields from three binary stellar evolution channels are available: channels A, B, and C. These channels could be, for example, the previously discussed case of mass transfer from an evolved donor to a main sequence companion, merged stars that merged on the main sequence, and merged stars that merged after the main sequence. These yields could then be included in the effective stellar yield calculation:

\begin{equation}
    Y_{\rm eff} = \frac{(1-h)Y_{\rm single} + h(f_{\rm A} Y_{\rm A} + f_{\rm B} Y_{\rm B} + f_{\rm C} Y_{\rm C})}{1 + h\langle q \rangle}.
    \label{eq:yeffabc}
\end{equation}

Here $f$ is the fraction of binaries with the appropriate primary mass that undergoes the relevant channel and Y is the yield of a binary from that channel. 

In some cases, these fractions can be estimated analytically. Suppose a grid of stellar models computed at different orbital separations similar that computed by \cite{brinkman2023} is being used. An initial orbital period distribution (e.g. flat in $\log(P)$, following Opik's law) can be assumed and each model in the set can be assigned as representative of a region of that distribution. Then the weighting of each model is simply the birth probability distribution integrated over the region for which the stellar model is assumed representative.

For more complicated evolution channels, such as stellar mergers, the $f$ can be obtained from binary population synthesis calculations. Recall that, just as in equation \ref{eq:brink2019yeff}, the effective stellar yield calculation is computed for each primary mass. There is, therefore, no restriction to mass-independence for any of the weighting factors that control the mix, such as the binary fraction $h$ or the evolution channel fractions $f$, or even the assumed distribution of $q$. Further, a more comprehensive binary population synthesis study could self-consistently provide these fractions not only as a function of the initial primary mass but also as a function of metallicity and binary stellar physics assumptions. Such a wealth of information could then be propagated into the GCE calculations through the effective stellar yield framework, providing feedback back to the binary stellar physics community. Another approach would be to treat the fractions as free parameters, and solve for them numerically by statistically comparing the resulting GCE calculations to observations of stellar abundances.

The single yield term can also be subdivided to better capture diversity in single stellar physics, which in nature extends beyond their initial mass. Suppose that a modeller wishes to employ a mixed population of rotating and non-rotating (or magnetic and non-magnetic) single star models, and several different rotations are available. Their inclusion can be done following the method for binaries: subdivide the single term, and then assign (possibly mass-dependent) fractions to each model-set to build your mixed stellar population. 

In this way, the calculation of the effective yield can become arbitrarily complex (and complete), inclusive of a wide range of different stellar yields from different evolution channels. Both approaches for obtaining the fractions that determine the weightings for the different evolution channels -- drawing them from binary population synthesis or solving for them as free parameters --  have the potential to provide valuable feedback to evolutionary modellers. By including a more complete picture of stellar evolution in our galactic chemical evolution calculations, we can better understand the limits of our current knowledge and make better use of our observational data.

\begin{table}[t]
    \centering
    \caption{Summary of Composition Models for Accreted Material}
    \renewcommand{\arraystretch}{1.0}
    \begin{tabular}{|p{2.5cm}|p{5.5cm}|}
        \hline
        \textbf{Model Name} & \textbf{Description} \\ \hline
        CMCL1 & Ignores composition of accreted material; only considers the effect of increasing the secondary mass on the main sequence. \\ \hline
        CMCL2 & Preserves net yields from the primary by modifying secondary yields based on enhancements/depletion in the accreted material. \\ \hline
        CMCL3 & Special treatment for fragile isotopes (Li, Be, B) to approximate the nuclear processing that would occur during the evolution of the secondary anyway. \\ \hline
        CMCL4 & Includes radioactive decay of accreted radioisotopes during the secondary’s remaining lifetime. \\ \hline
    \end{tabular}
    \label{tab:composition_models}
\end{table}

\begin{figure}
\centering
\begin{subfigure}{0.99\columnwidth}
\centering
\includegraphics[width=0.99\columnwidth]{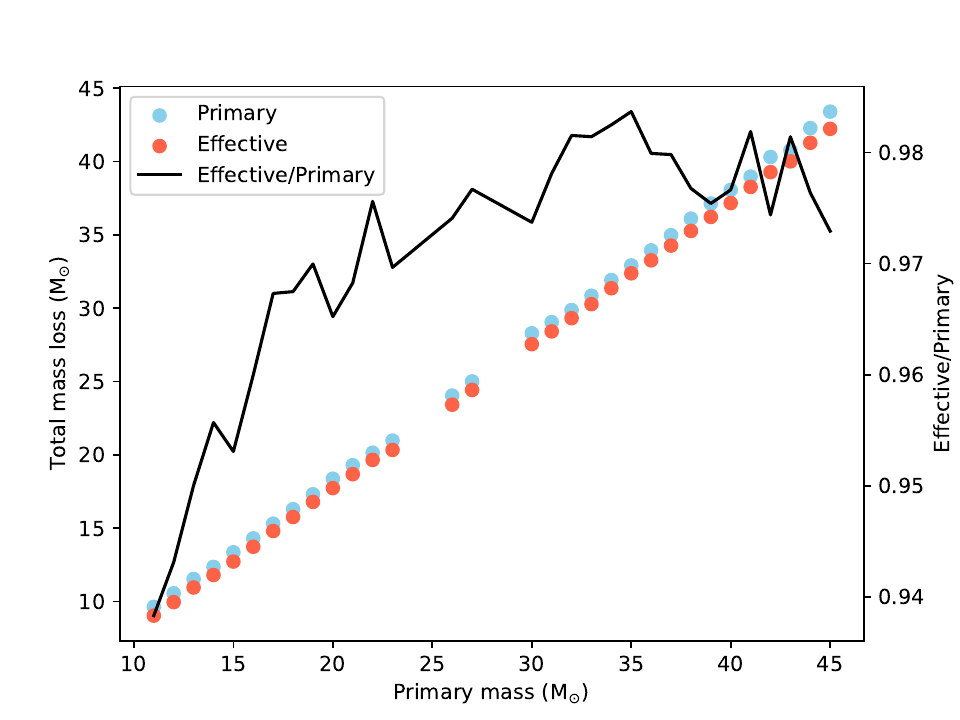}
\end{subfigure}

\begin{subfigure}{0.99\columnwidth}
\centering
\includegraphics[width=0.99\columnwidth]{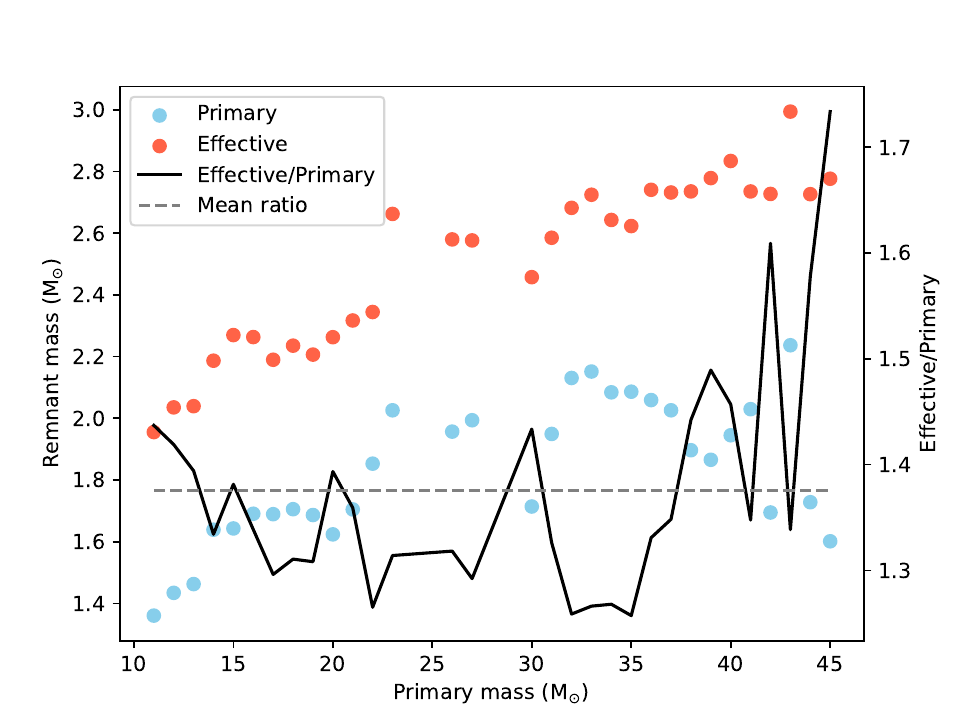}
\end{subfigure}

\begin{subfigure}{0.99\columnwidth}
\centering
\includegraphics[width=0.99\columnwidth]{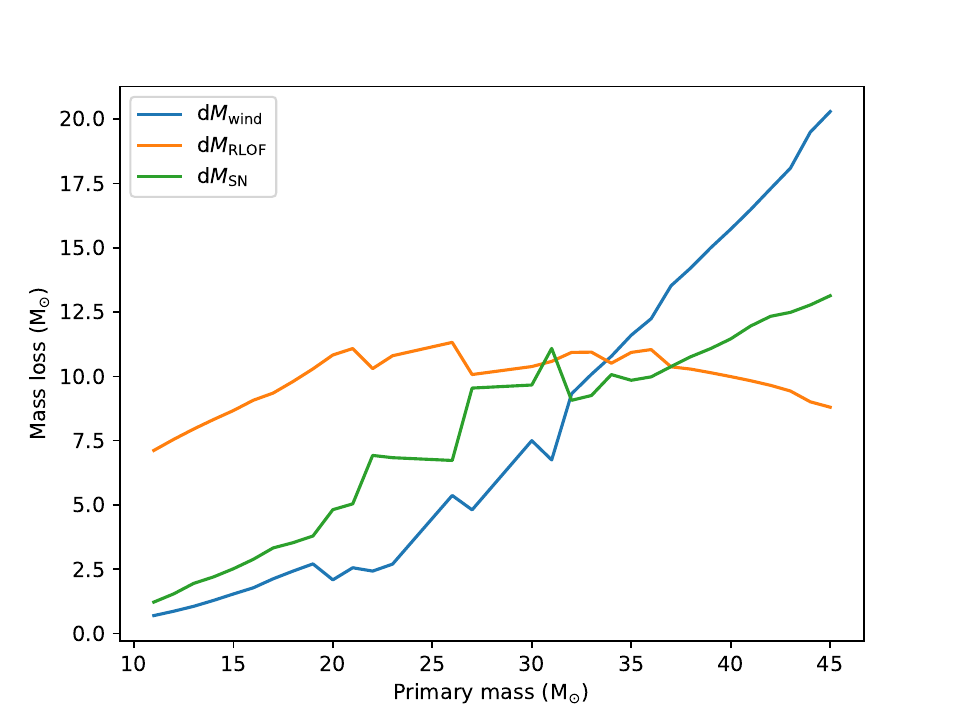}
\end{subfigure}%

\caption{Mass budget of the \cite{farmer2023} model primaries compared to an effective stellar mass budget ($h=0.5$, $\beta_{\rm winds}=0.1$, $\beta_{\rm RLOF}=1$, and $\beta_{\rm SN}=0$). The upper panel shows the total mass loss vs the primary mass for both the primary yields reported in \cite{farmer2023} and the effective stellar yield. The middle panel shows the implied mass locked away in stellar remnants in the reported vs effective stellar yield calculation. The lower panel breaks down the amount of mass lost through each mechanism (see also Tables 3-6 in \cite{farmer2023}.}
\label{fig:mass_loss}
\end{figure}

\section{Results}
\label{sec:res}

In this section, we describe the impact of a range of different physical assumptions on the effective stellar yields. First, we  consider the effect of modelling the composition of the accreted material on the effective stellar yields, while fixing the binary fraction and all mass transfer efficiencies ($\beta_{\rm winds}$, $\beta_{\rm RLOF}$, and $\beta_{\rm SN}$). We then explore the effect of variable binary physics through the binary fraction and binary accretion efficiency parameters.

Equivalent figures to those described in this section are available in the online supplementary material for all isotopes computed in \cite{farmer2023}, along with pre-computed tables of effective stellar yields for a wide range of binary parameter assumptions.

\subsection{Composition of accreted material}
\label{sec:res:comp}

The simplest treatment for the composition of accreted material is to ignore it, and instead only consider the effect of increasing the secondary mass on the main sequence. We  refer to this model as Composition-Model-Complexity-Level-1, or CMCL1.

The next step up in complexity is a model where all net yields from the primary are preserved by the secondary (CMCL2). That is, the secondary's yields are modified according to any net enhancement or depletion in the accreted material. Note that the mass-increase effect is also accounted for; the difference with CMCL1 is that the chemical signature of the primary's evolution history remains. In reality, secondary processing on the enriched accreted material will likely occur at some level, but a full accounting of this secondary processing requires detailed stellar model calculations.

For the fragile light elements Li, Be, and B, however, a simple treatment that assumes their nuclear processing and destruction but prevents duplicate destruction will likely suffice. In CMCL3, we introduce a special treatment for each isotope of the fragile, light elements Li, Be, and B in an attempt to increase the fidelity for these isotopes and their daughter products (see Section \ref{sec:effective_yields}).

Building on CMCL3, we introduce one more improvement to our treatment of the accreted material: accounting for the radioactive decay of accreted radioisotopes during the remaining lifetime of the secondary (CMCL4). This is the highest fidelity composition model we consider in this work. Details on how we treat the decay process can be found in Section \ref{sec:effective_yields}, and a summary of all considered treatments for the composition of the accreted material can be found in Table \ref{tab:composition_models}.

We consider these four composition models for a fixed set of binary assumptions: $h=0.5$, $\beta_{\rm winds}=0.1$, $\beta_{\rm RLOF}=1$, and $\beta_{\rm SN}=0.05$. We adopt a high value of $\beta_{\rm SN}=0.05$ to illustrate the effect of altering the composition model on isotopes formed only during the supernova explosion. In nature, it is unlikely that 5\% of the supernova ejecta would be accreted onto the companion.

\begin{figure*}
\centering
\begin{subfigure}{0.99\columnwidth}
\centering
\includegraphics[width=0.99\columnwidth]{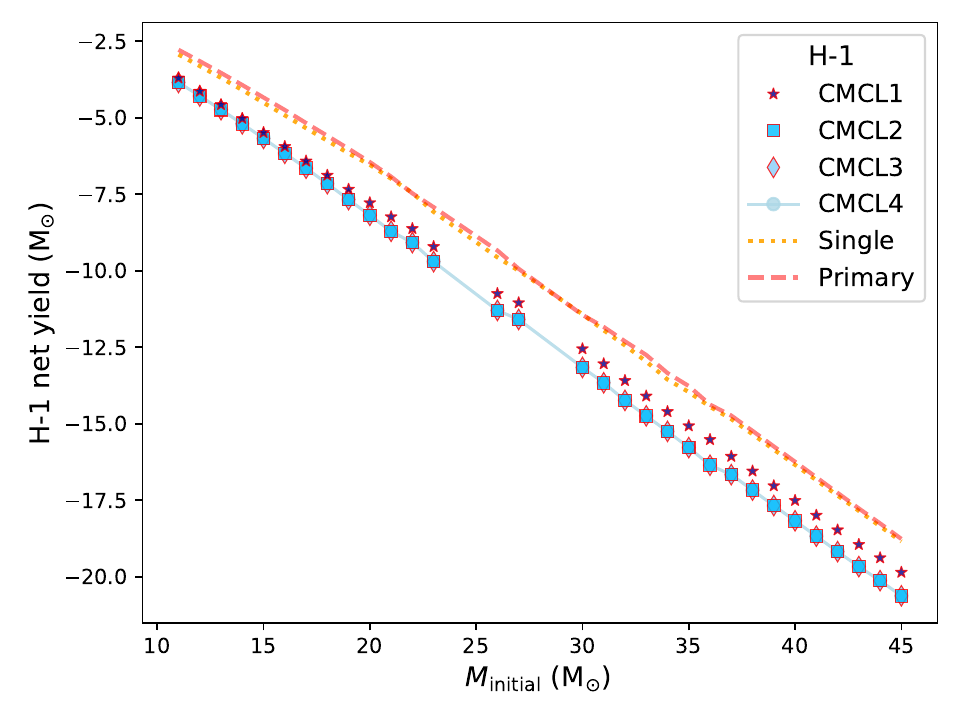}
\end{subfigure}%
\begin{subfigure}{0.99\columnwidth}
\centering
\includegraphics[width=0.99\columnwidth]{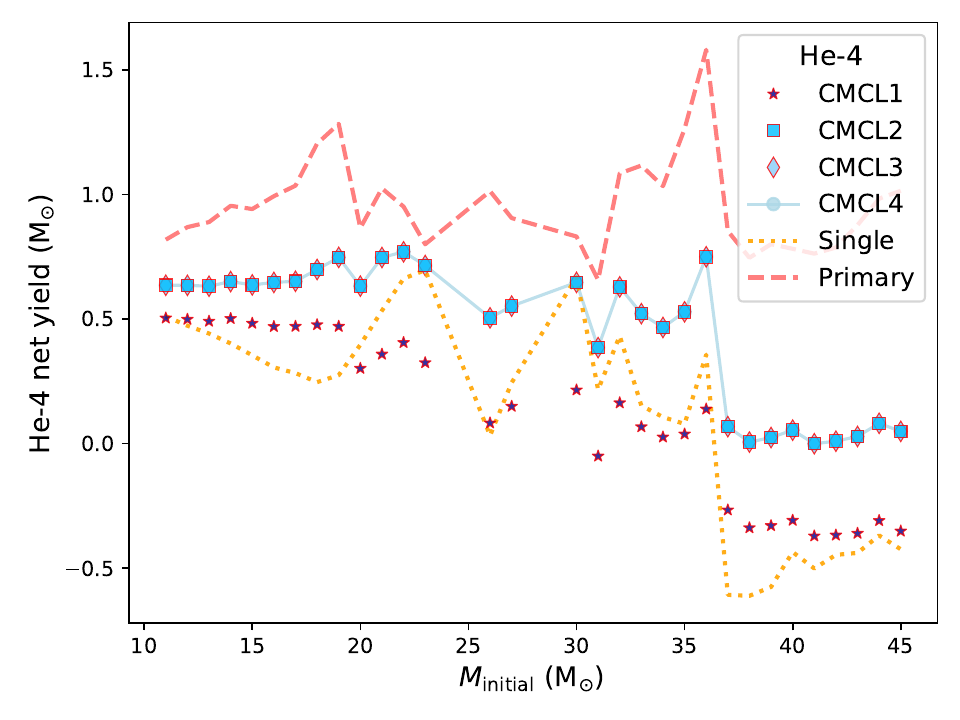}
\end{subfigure}

\begin{subfigure}{0.95\columnwidth}
\centering
\includegraphics[width=0.95\columnwidth]{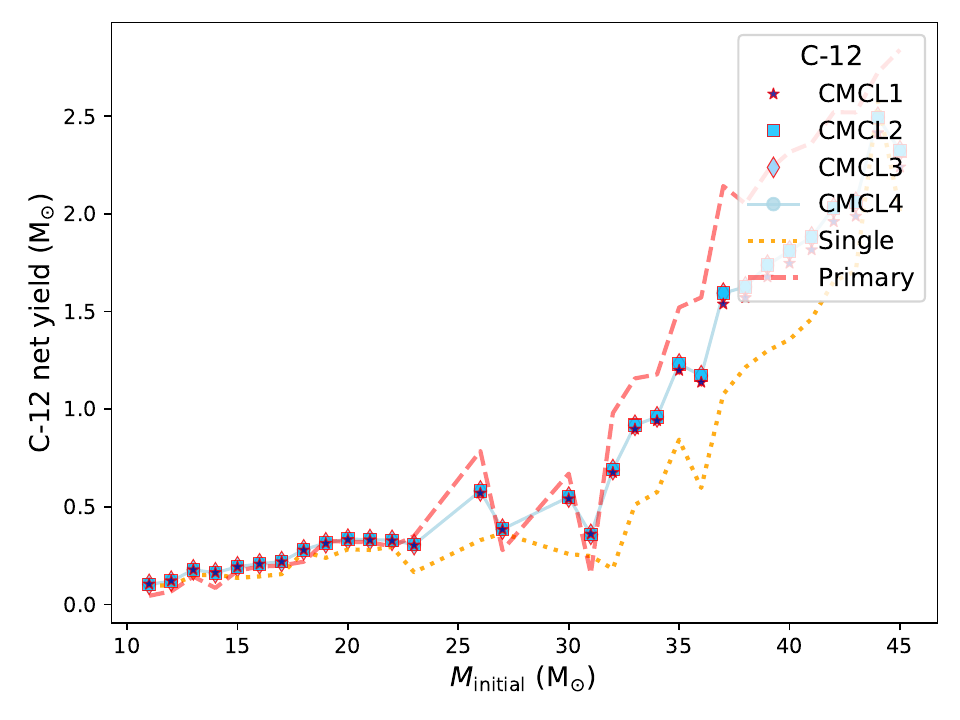}
\end{subfigure}%
\begin{subfigure}{0.99\columnwidth}
\centering
\includegraphics[width=0.99\columnwidth]{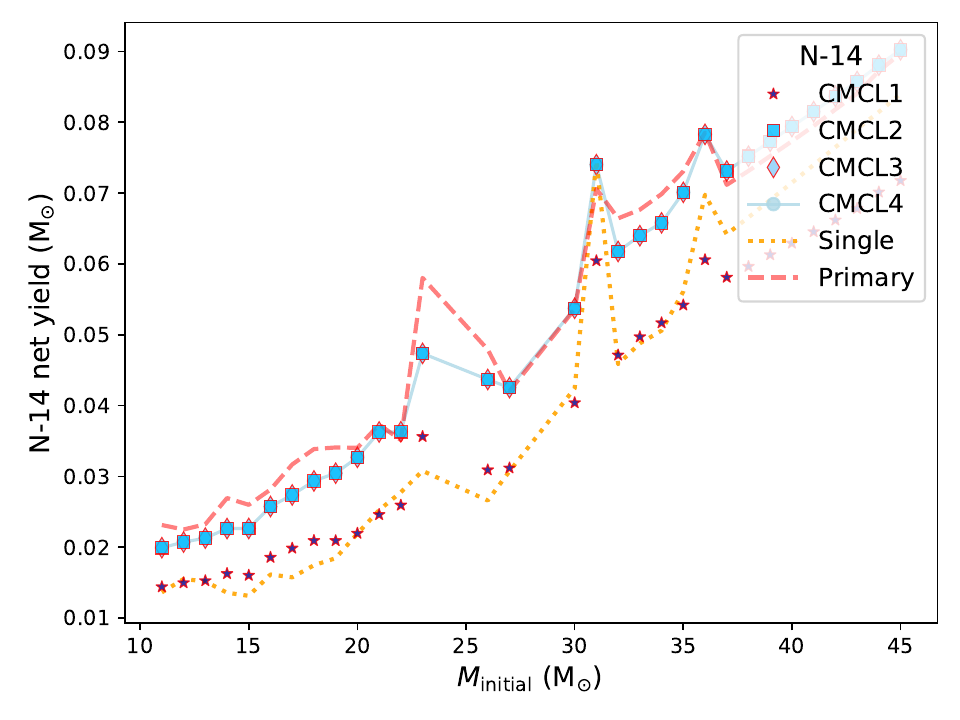}
\end{subfigure}

\begin{subfigure}{0.99\columnwidth}
\centering
\includegraphics[width=0.99\columnwidth]{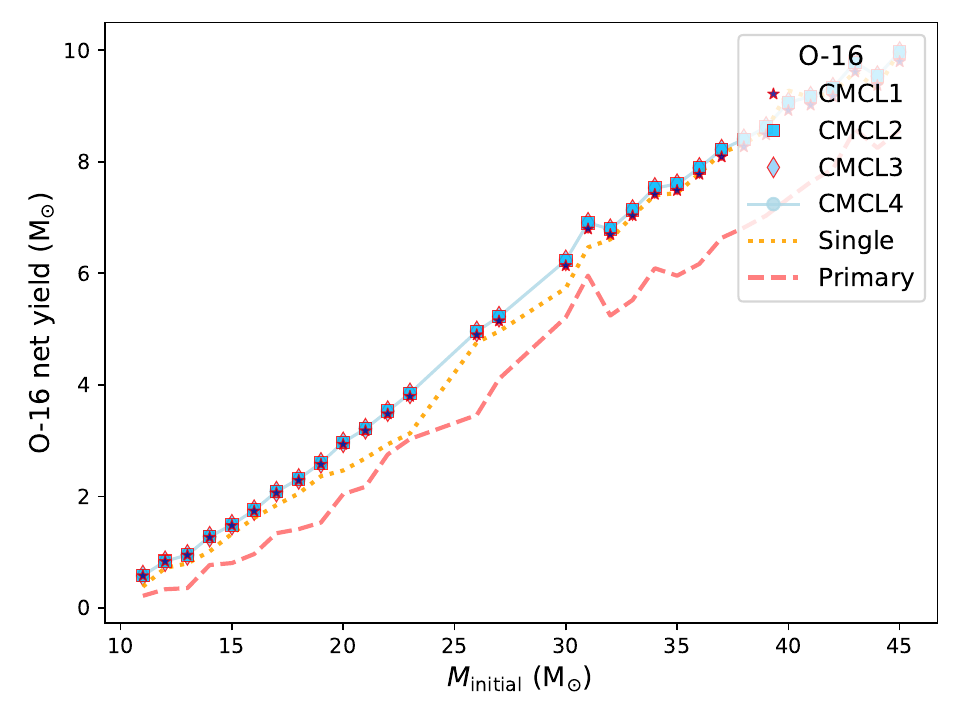}
\end{subfigure}%
\begin{subfigure}{0.99\columnwidth}
\centering
\includegraphics[width=0.99\columnwidth]{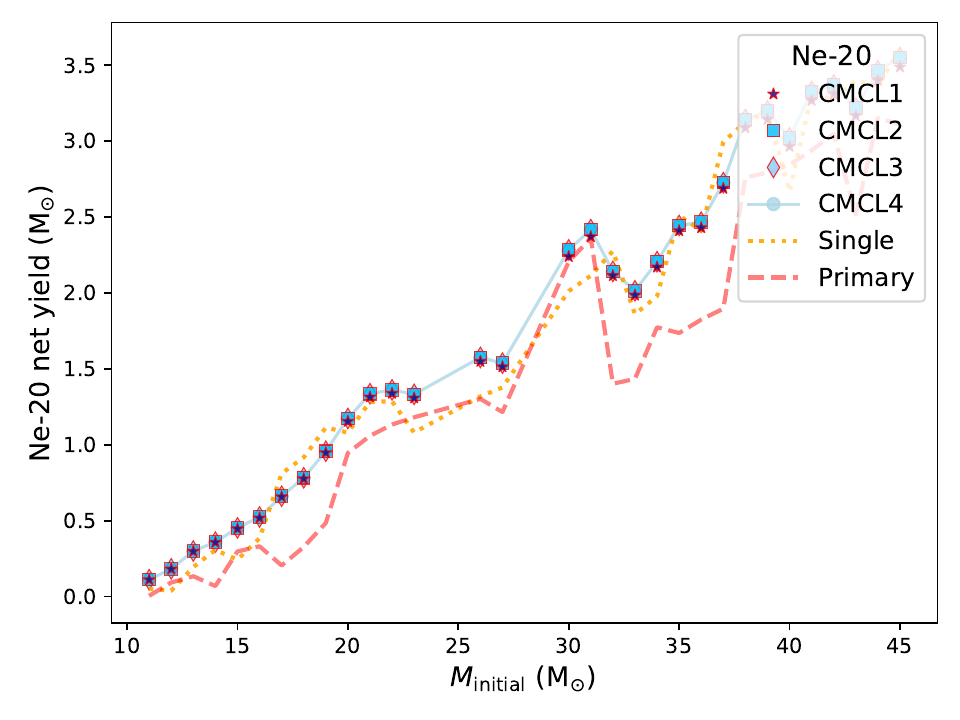}
\end{subfigure}
\caption{Net yields for the dominant isotopes of H, He, C, N, O, and Ne. Effective yields computed for $h=0.5$, $\beta_{\rm winds}=0.1$, $\beta_{\rm RLOF}=1$, and $\beta_{\rm SN}=0.05$ are shown in shades of blue. The dashed lines present the net yields of each isotope for the reported single star (orange) and the reported primary star (red) for comparison.}
\label{fig:cmcl_HHeCNO}
\end{figure*}

Fig. \ref{fig:cmcl_HHeCNO} presents effective net yields for the dominant isotopes of H, He, C, N, O, and Ne for the different composition models. The reported single and primary yields from \cite{farmer2023} are over-plotted for comparison. The first thing to note is that the treatment of the composition model affects even the net H yield, with CMCL1 showing higher (less negative) net H yields because it fails to account for H depletion in the accreted material. CMCL2, CMCL3, and CMCL4 are consistent with each other for H, as we would expect given that they all preserve the primary's net H yield.

We can also see that all plotted effective yields, even CMCL1, differ from the reported yields in \cite{farmer2023}. This is partly because the total mass ejected is slightly different in the binary scenario (see top panel of Fig. \ref{fig:mass_loss}), as the effective remnant mass that is `locked away' is larger than the remnant mass of the primary (see middle panel of Fig. \ref{fig:mass_loss}). This is the statistical consequence of accounting for the secondary's (usually higher) remnant-to-initial mass ratio in the effective yield calculation. This reduces the effective mass lost from the binary, the effect of which is most easily seen in the H yield.

The effective net yields of He and N are significantly lower under the CMCL1 composition model compared to the models accounting for the composition of the accreted material (CMCL2, CMCL3, CMCL4), while C, O, and Ne are barely affected by the composition model. This is intuitive, as most of the mass transferred under this set of accretion efficiency assumptions is through RLOF.  Hence, the elements that the envelope of the primary is most enriched in -- He and N -- will be most affected by the whether or not the primary's nuclear processing is preserved.

\begin{figure*}
\centering
\begin{subfigure}{0.99\columnwidth}
\centering
\includegraphics[width=0.99\columnwidth]{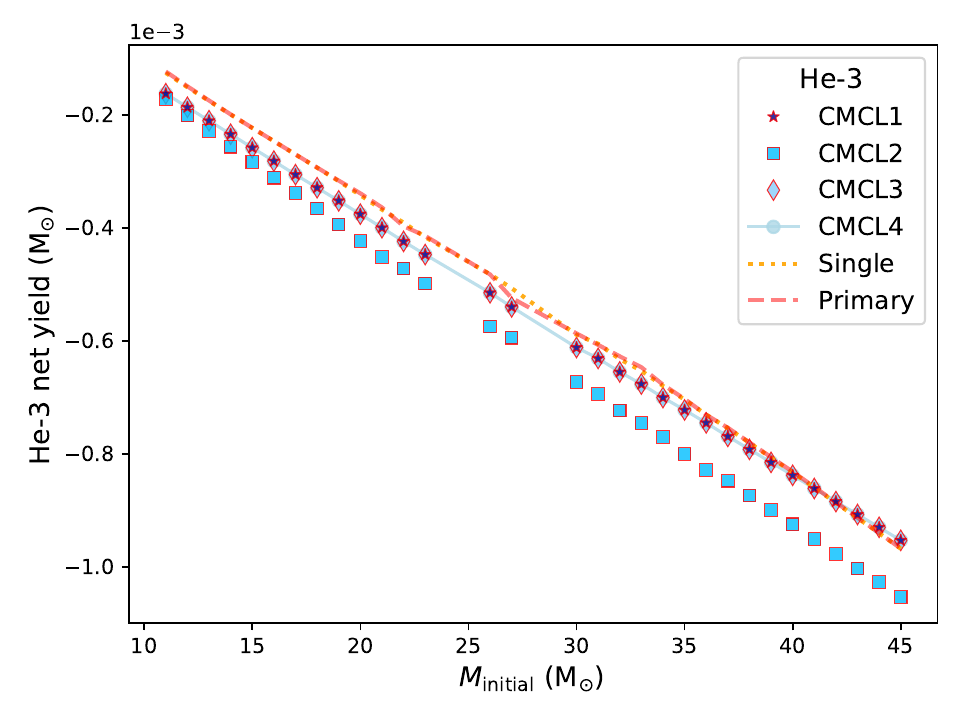}
\end{subfigure}%
\begin{subfigure}{0.99\columnwidth}
\centering
\includegraphics[width=0.99\columnwidth]{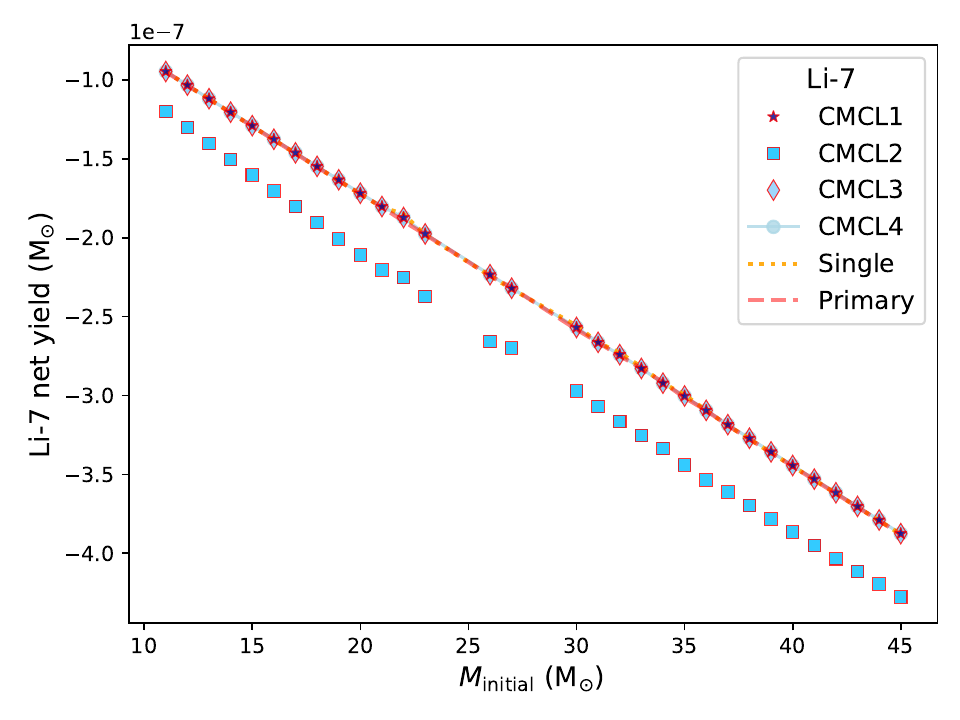}
\end{subfigure}

\begin{subfigure}{0.99\columnwidth}
\centering
\includegraphics[width=0.95\columnwidth]{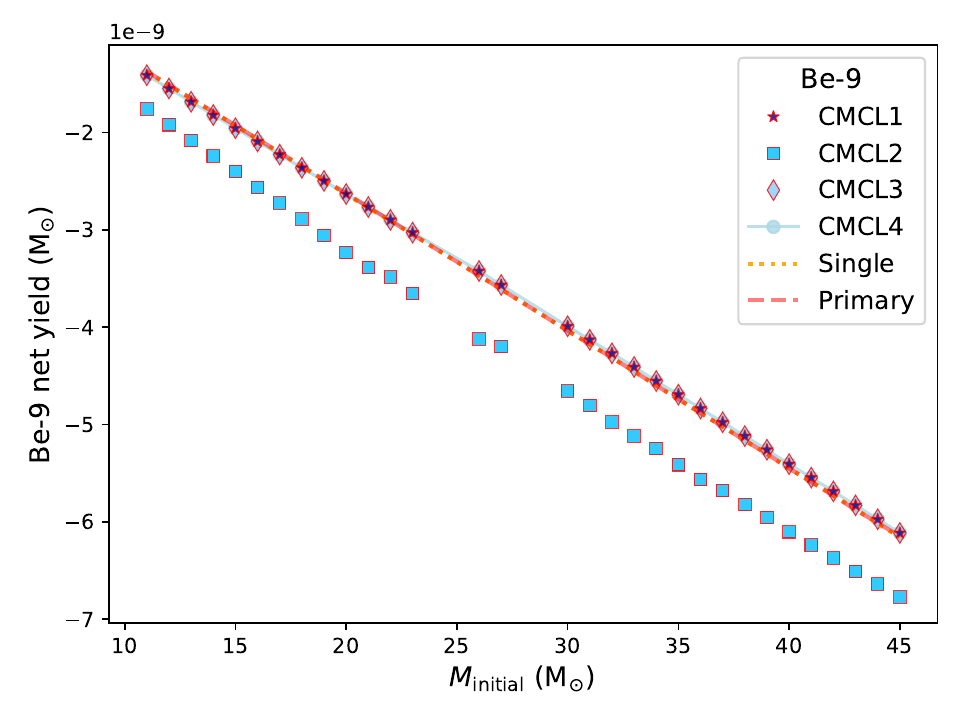}
\end{subfigure}%
\begin{subfigure}{0.99\columnwidth}
\centering
\includegraphics[width=0.95\columnwidth]{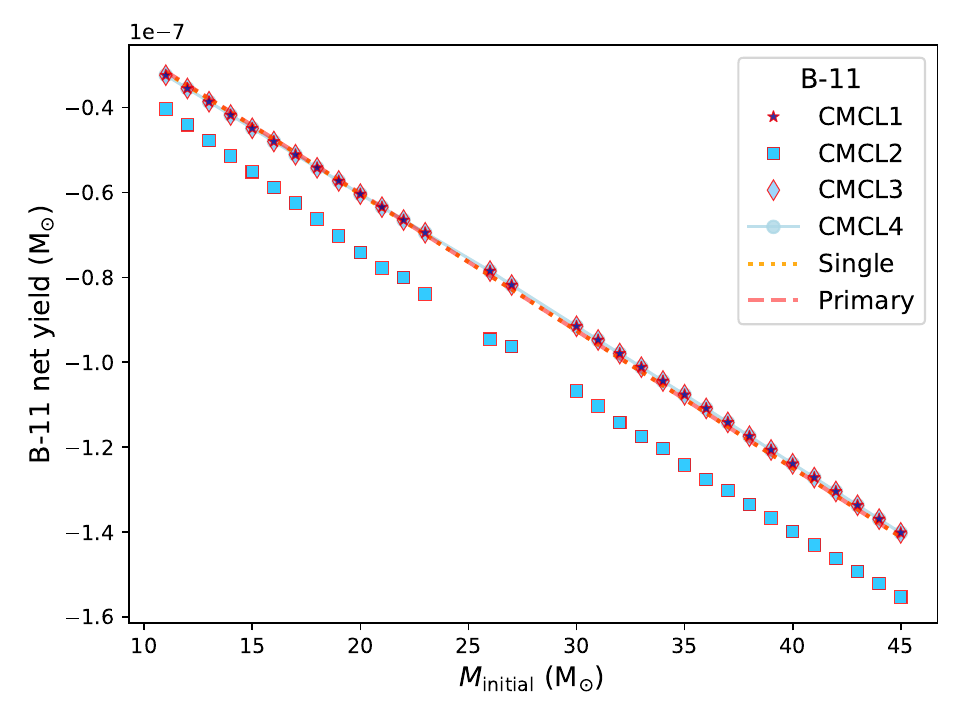}
\end{subfigure}

\caption{Net yields for He-3, Li-7, Be-9, and B-11. Effective yields computed for $h=0.5$, $\beta_{\rm winds}=0.1$, $\beta_{\rm RLOF}=1$, and $\beta_{\rm SN}=0.05$ are shown in shades of blue. The dashed lines present the net yields of each isotope for the reported single star (orange) and the reported primary star (red) for comparison.}
\label{fig:cmcl_light}
\end{figure*}

Fig. \ref{fig:cmcl_light} shows the effective net yields of the light isotopes He-3, Li-7, Be-9 and B-11. CMCL2, which preserves the processing of the primary but does not include our artificial correction for the nuclear processing of accreted light elements in the secondary, is clearly discrepant with the other composition models, showing lower (more negative) yields of He-3, Li-7, and B-11 but higher Be-7. All other composition models are consistent, as CMCL3 and CMCL4 both include the same treatment for the light elements, while CMCL1 assumes accretion of un-processed material, which also has the effect of erasing any net (positive or negative) yields in these isotopes caused by nuclear processing in the primary.

\begin{figure*}
\centering
\begin{subfigure}{0.99\columnwidth}
\centering
\includegraphics[width=0.99\columnwidth]{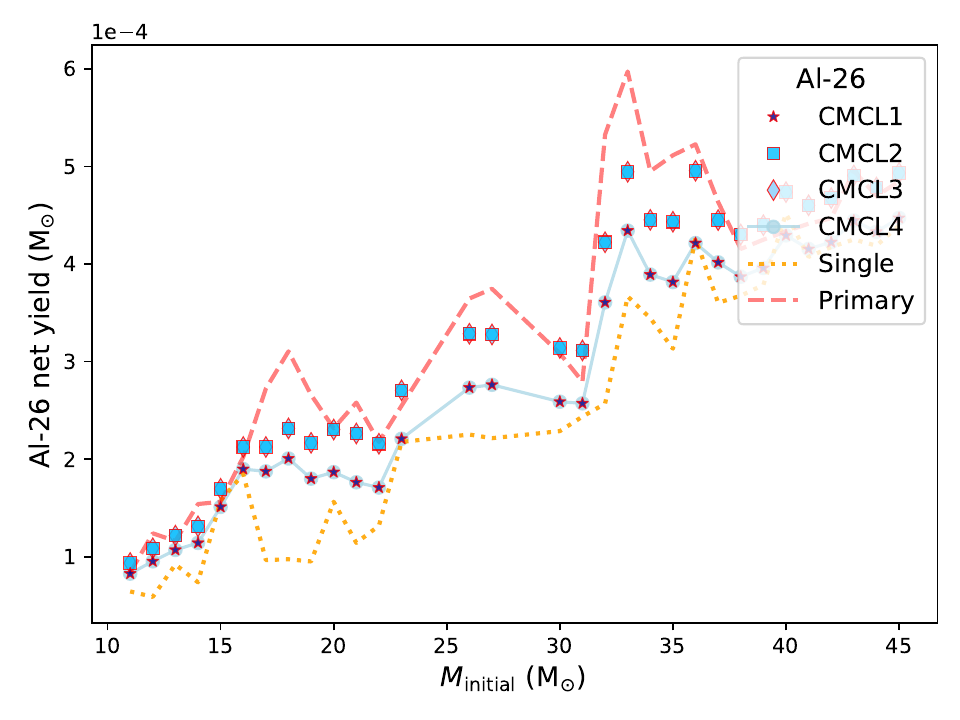}
\end{subfigure}%
\begin{subfigure}{0.99\columnwidth}
\centering
\includegraphics[width=0.99\columnwidth]{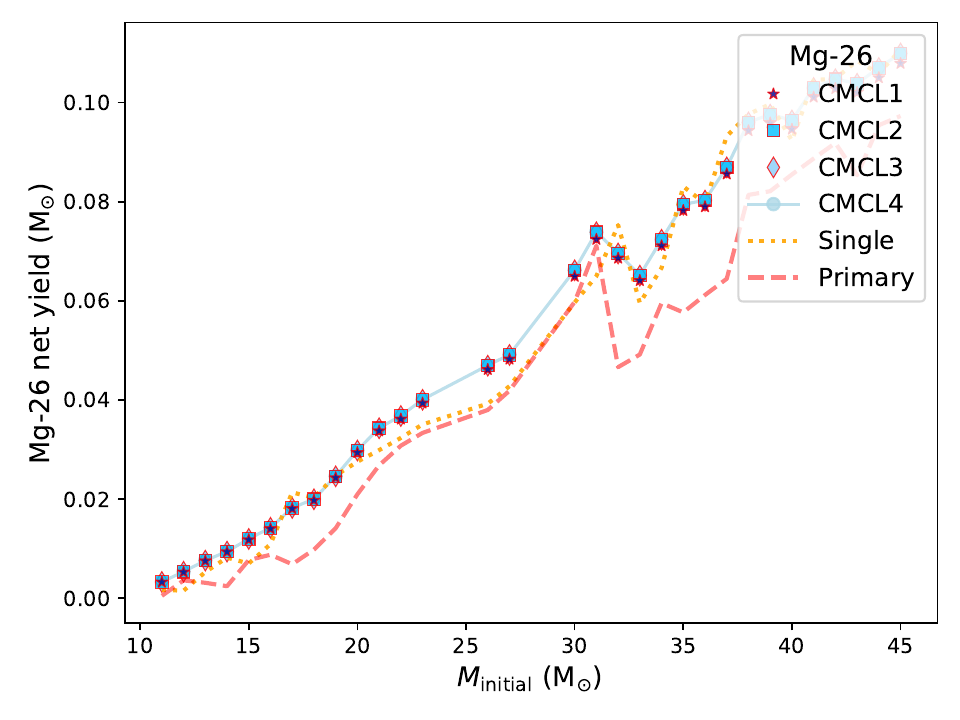}
\end{subfigure}

\begin{subfigure}{0.99\columnwidth}
\centering
\includegraphics[width=0.99\columnwidth]{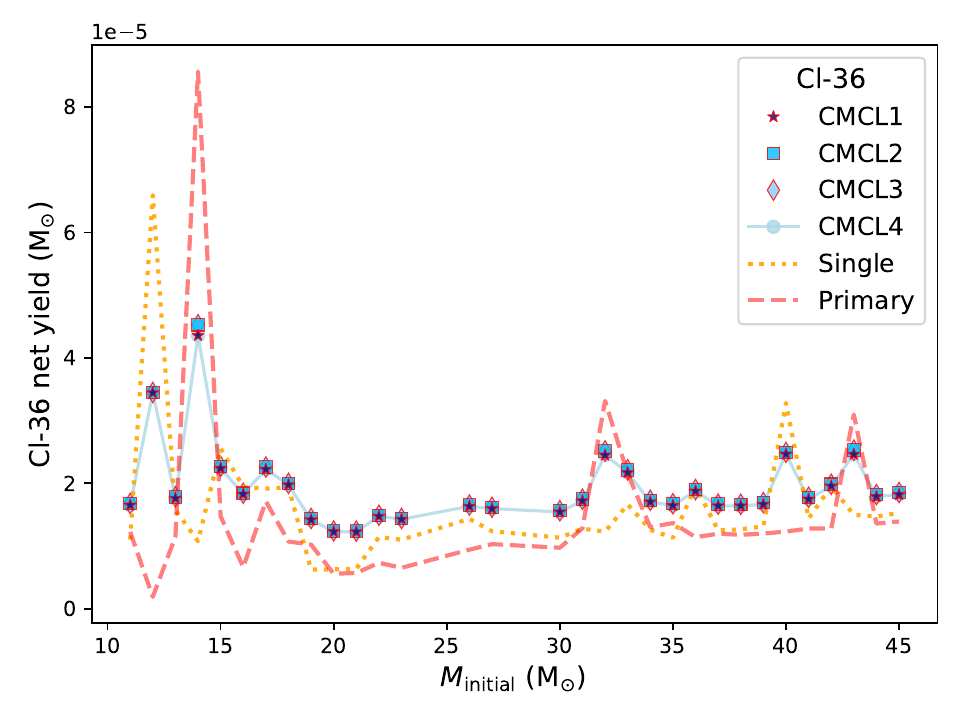}
\end{subfigure}%
\begin{subfigure}{0.99\columnwidth}
\centering
\includegraphics[width=0.99\columnwidth]{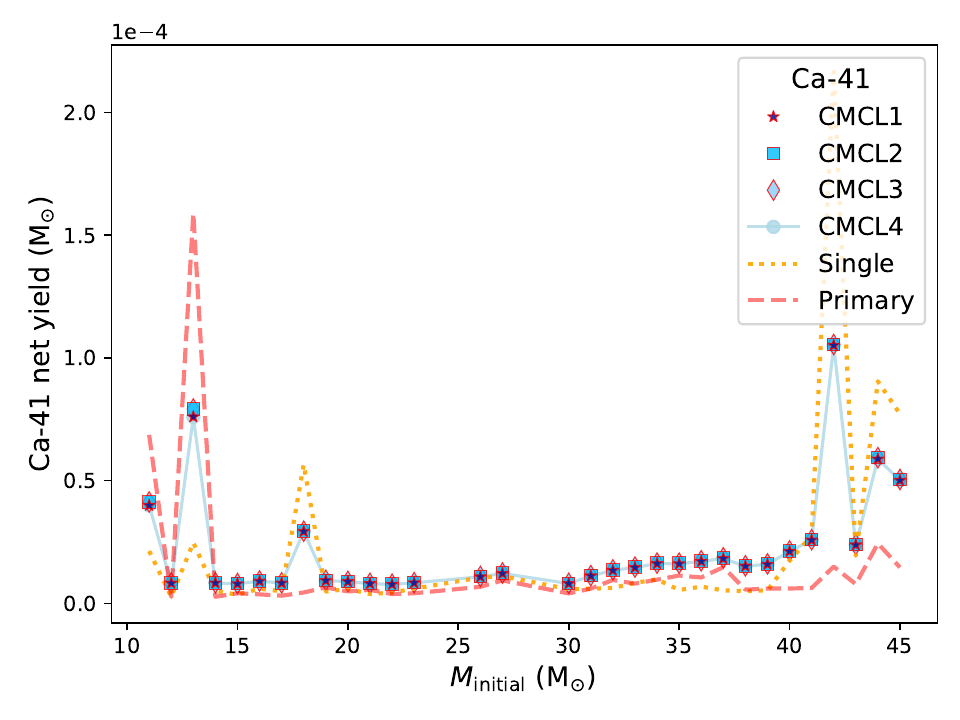}
\end{subfigure}

\begin{subfigure}{0.99\columnwidth}
\centering
\includegraphics[width=0.99\columnwidth]{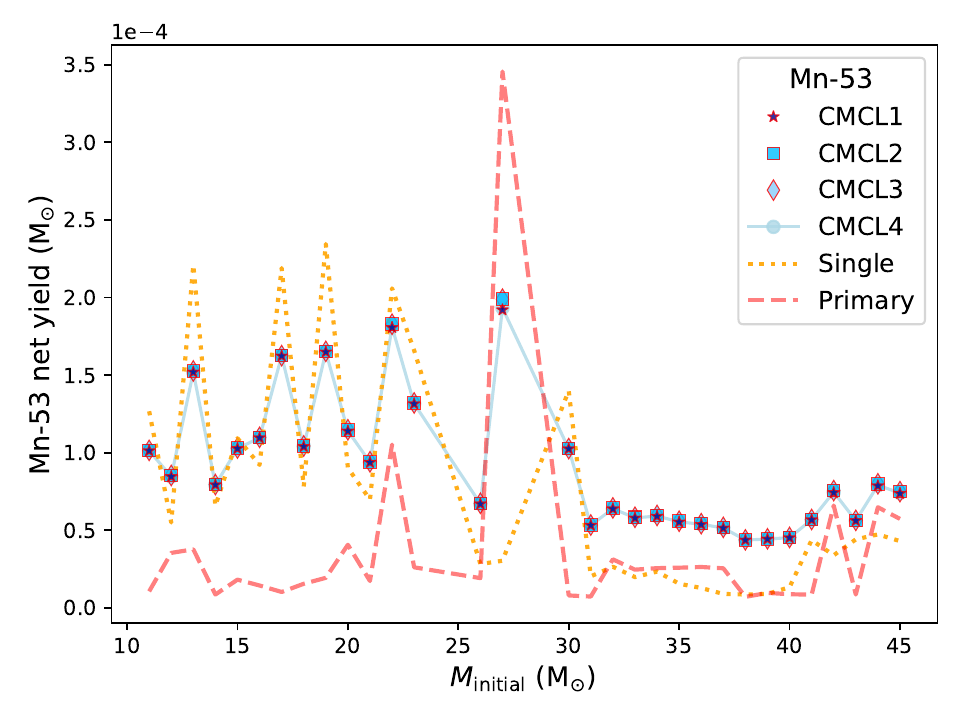}
\end{subfigure}%
\begin{subfigure}{0.99\columnwidth}
\centering
\includegraphics[width=0.99\columnwidth]{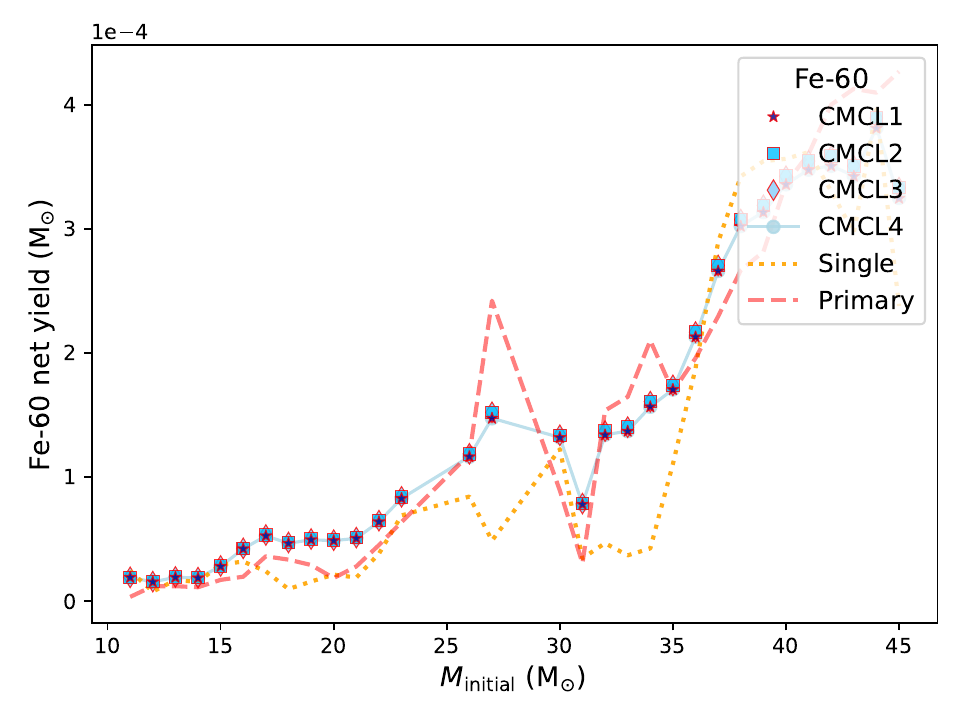}
\end{subfigure}

\caption{Net yields for the dominant isotopes of Al-26, Mg-26, Cl-36, Ca-41, Mn-53, and Fe-60. Effective yields computed for $h=0.5$, $\beta_{\rm winds}=0.1$, $\beta_{\rm RLOF}=1$, and $\beta_{\rm SN}=0.05$ are shown in shades of blue. The dashed lines present the net yields of each isotope for the reported single star (orange) and the reported primary star (red) for comparison.}
\label{fig:cmcl_radio}
\end{figure*}

Fig. \ref{fig:cmcl_radio} shows the effective yields for the radioisotopes Al-26, Cl-36, Ca-41, Mn-53, and Fe-60, as well as Al-26's daughter nuclide Mg-26. The yields of Al-26 from \cite{farmer2023} are known to overestimate Al-26 yields due to not including a short-lived ($\approx6$~s half-life) isomeric state of Al-26, significantly affecting their Al-26 yields. Despite this caveat, the Al-26 panel is still worth briefly discussing, as it is one of the few short-lived radioisotopes that can be produced during hydrostatic burning in significant quantities.

The composition models accounting for the nuclear processing for the primary, but not radioactive decays during the post-mass transfer life-time of the secondary (CMCL2 and CMCL3), are strongly discrepant from CMCL1 (where the composition of the ejecta is not accounted for) and CMCL4, where its decay to Mg-26 is accounted for. The strong agreement between CMCL1 and CMCL4 reflects the fact that the half-life of Al-26 (\timestento{7}{5}~yr, \citealt{basunia2016}) is short relative to the remaining lifetime of the secondary for all but the closest mass-ratio binaries, allowing plenty of time for most of the Al-26 to decay. Thus, we can conclude that binary stars may actually diminish, rather than increase~\citep{brinkman2019,brinkman2023}, Al-26 yields originating from hydrostatic nuclear burning, as material that may otherwise have been lost to the ISM through stellar winds will instead decay in the envelope of the secondary. Regarding Al-26's daughter-product, Mg-26, we note that its effective yield is almost identical regardless of the Al-26 yield due to its alternate production channel during supernova explosions. The other radioisotopes considered do not experience significant production during hydrostatic burning, and so little is accreted onto the secondary. Thus the composition model makes little difference to the effective yield.

We adopted a value for the accretion efficiency of supernova ejecta of $\beta_{\rm SN}=0.05$, at the high end of what might be considered plausible. Even in this optimistic scenario, there is insufficient material accreted by the secondary for the treatment of its composition to have a noticeable effect on the effective yield of isotopes produced during supernovae. The effect of changing the accretion efficiencies is dealt with in detail in Section \ref{sec:res:binfrac}, but for now we simply note that it takes an increase in the accretion efficiency of supernova material to 30\% is sufficient for significant deviations between the composition models to become apparent for Cl-36, Ca-41, and Fe-60, as shown in Fig. \ref{fig:cmcl_radio_cc30}. We note that for long-lived $(\gtrapprox$\tento{8}~yr) radioisotopes the importance of accounting for decay on the secondary will increase as the half-life becomes more comparable to the post-mass transfer evolutionary timescale of the secondary.

We conclude that under reasonable binary assumptions, the treatment of the composition of accreted material has a noticeable impact on isotopes for which there is significant enrichment or depletion (through hydrostatic burning and subsequent mixing) in the envelope's primary during RLOF, reflecting the fact that most of the accretion will occur during prior to explosive nuclear burning during the supernova. It is unlikely to impact isotopes which are only produced in significant quantities in SNe due to the low accretion efficiencies inherent to supernovae. We also acknowledge that all of the treatments described are merely first-order approximations of reality; some level of re-processing of accreted material will occur, an effect that may to be particularly impactful in low-metallicity environments where the relative composition change in the secondary due to accretion will be greater. A set of detailed models comparing the nucleosynthesis of post-accretion stars with single stars of comparable post-accretion mass would be of significant value in quantifying the error inherent to the described composition models.

\subsection{Binary parameters}
\label{sec:res:binfrac}

The effective stellar yields are computed as a function of the binary fraction, $h$, and the accretion efficiency parameters corresponding to different mass transfer channels: $\beta_{\rm winds}$, $\beta_{\rm RLOF}$, and $\beta_{\rm SN}$. In this section, we show the effect of varying each of these parameters, without strict adherence to physical realism, to provide some intuition about how these parameters can affect the effective stellar yields.

\subsubsection{Binary fraction $h$}

\begin{figure*}
\centering
\begin{subfigure}{0.99\columnwidth}
\centering
\includegraphics[width=0.99\columnwidth]{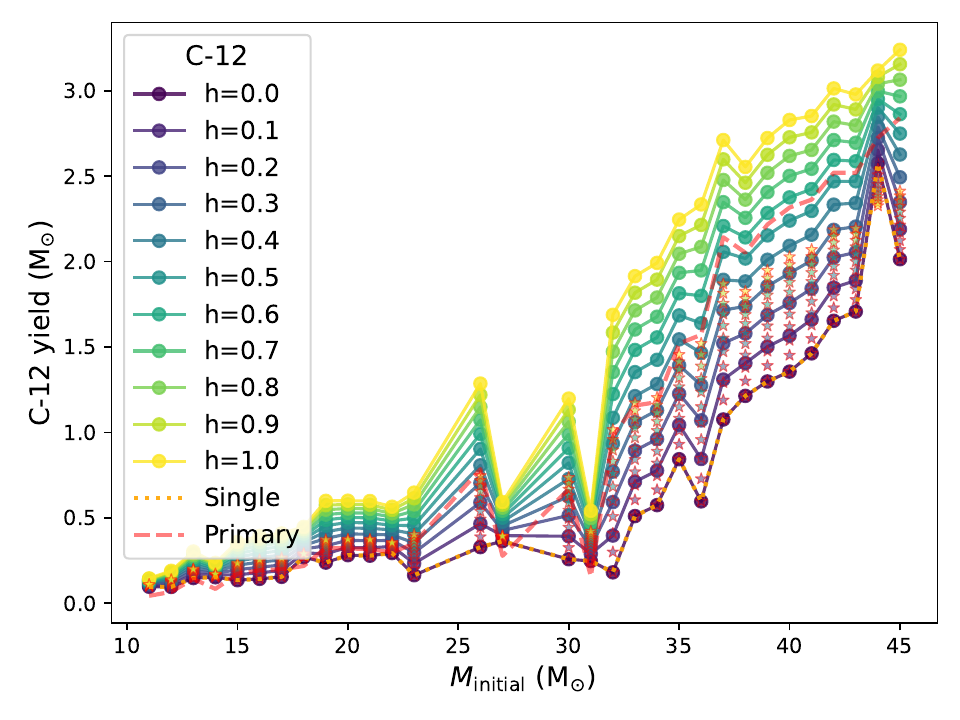}
\end{subfigure}%
\begin{subfigure}{0.99\columnwidth}
\centering
\includegraphics[width=0.99\columnwidth]{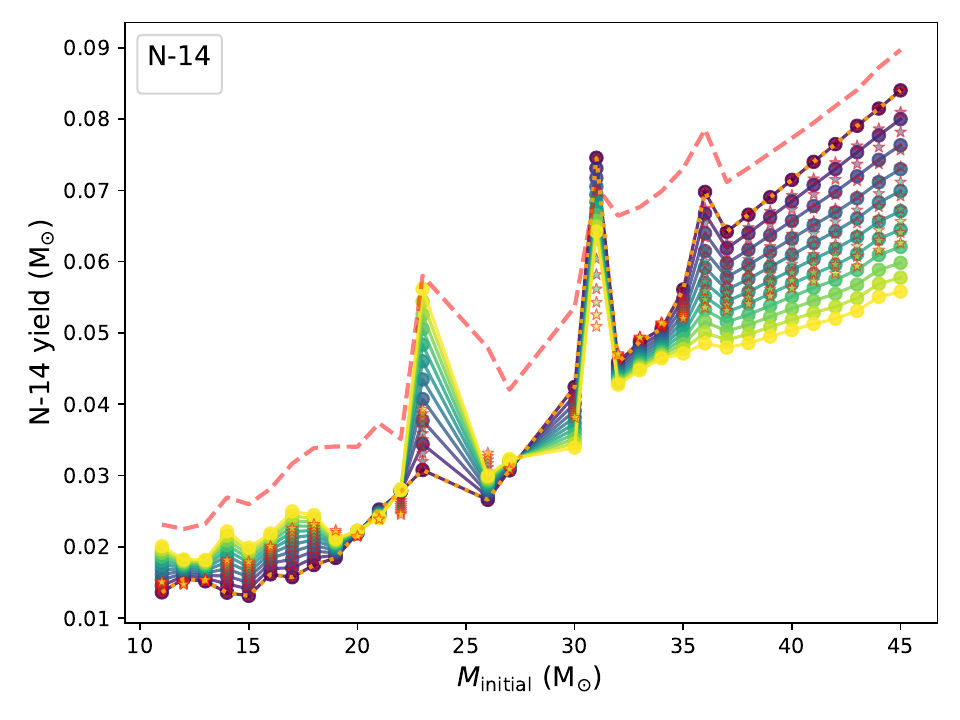}
\end{subfigure}

\begin{subfigure}{0.99\columnwidth}
\centering
\includegraphics[width=0.99\columnwidth]{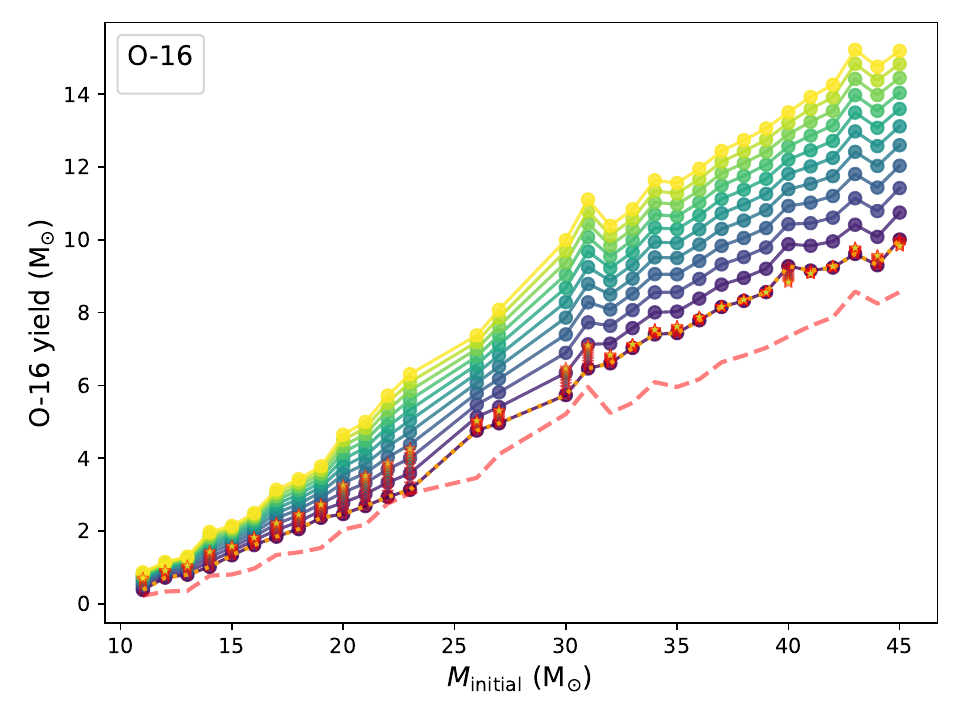}
\end{subfigure}%
\begin{subfigure}{0.99\columnwidth}
\centering
\includegraphics[width=0.99\columnwidth]{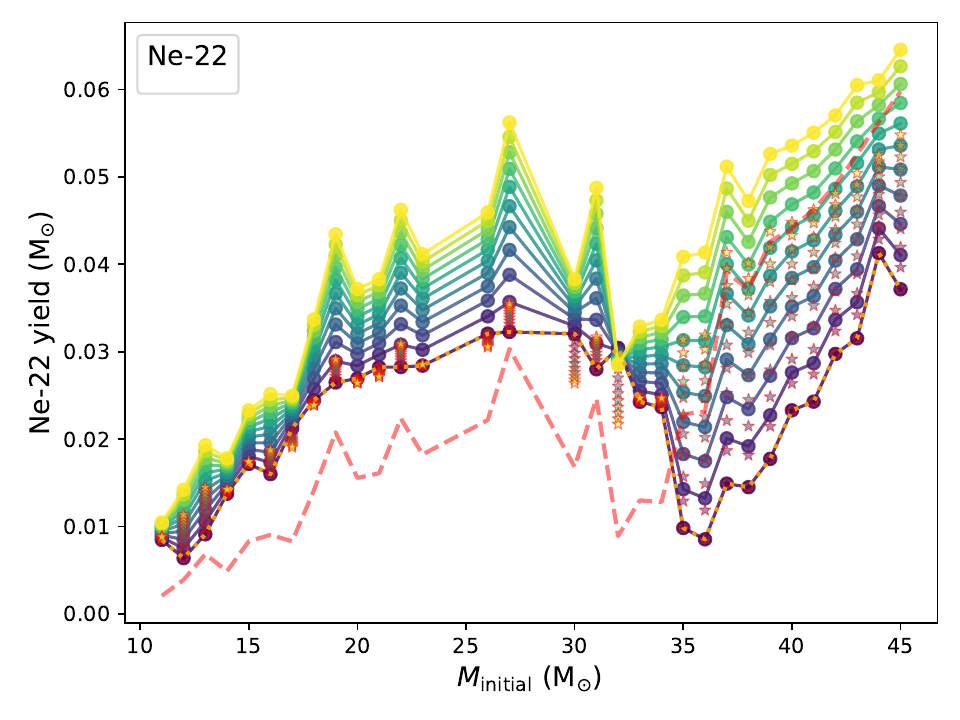}
\end{subfigure}

\begin{subfigure}{0.99\columnwidth}
\centering
\includegraphics[width=0.99\columnwidth]{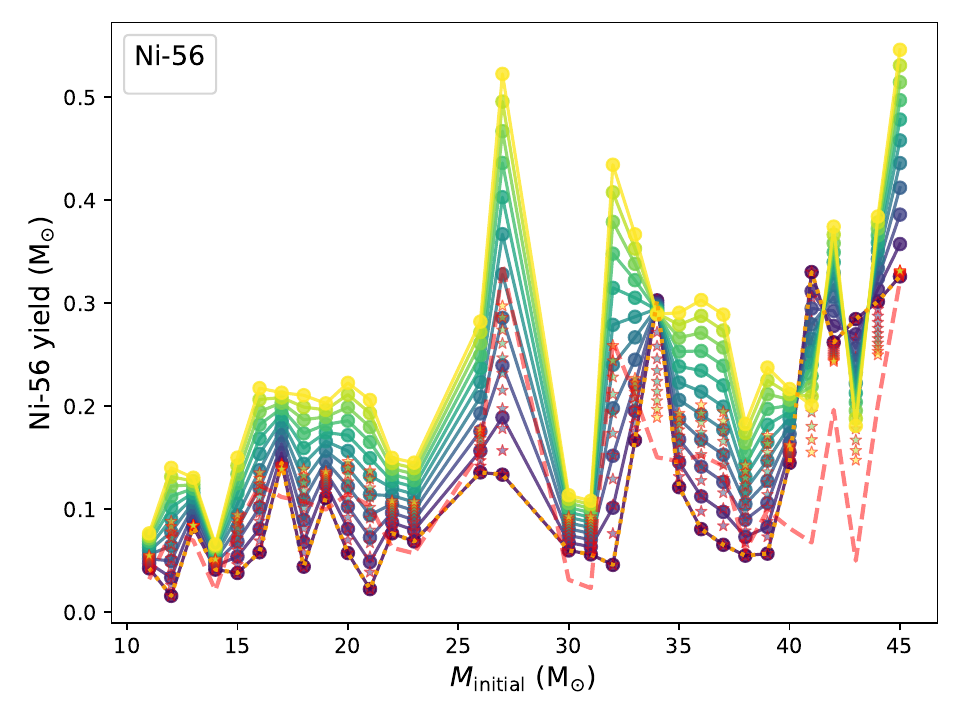}
\end{subfigure}%
\begin{subfigure}{0.99\columnwidth}
\centering
\includegraphics[width=0.99\columnwidth]{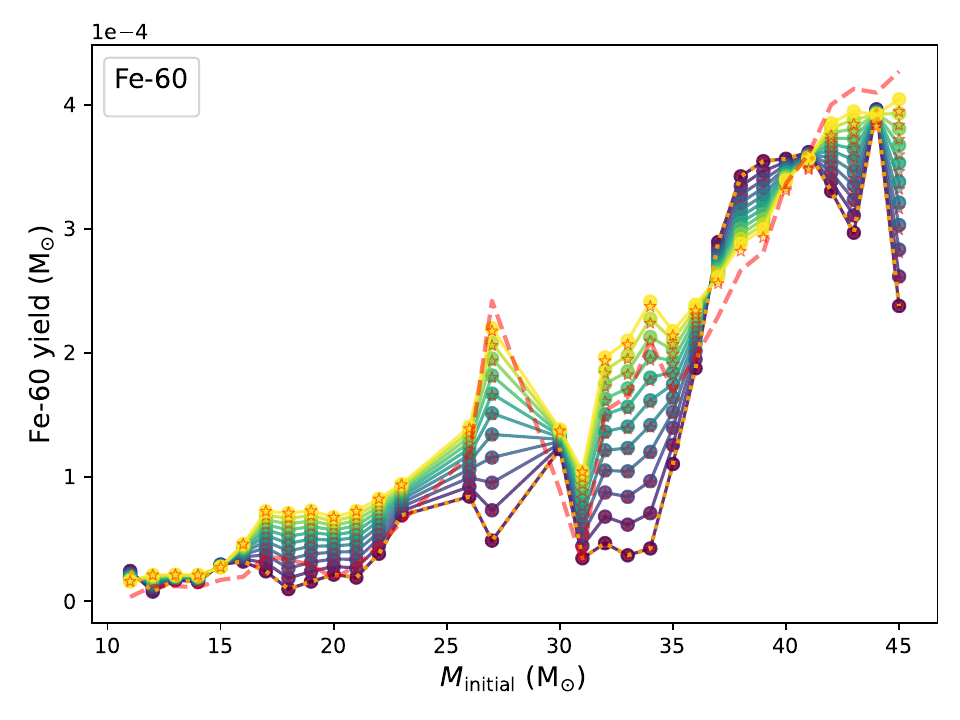}
\end{subfigure}

\caption{Effective net yields for C-12, N-14, O-16, Ne-22, Ni-56, and Fe-60 varying the binary fraction $h$ ($\beta_{\rm winds}=0.1$, $\beta_{\rm RLOF}=1$, and $\beta_{\rm SN}=0$). The connected dots were computed using CMCL4, while the isolated stars were computed using CMCL1. The dashed lines present the net yields of each isotope for the reported single star (orange, overlapping with the h=0 case) and the reported primary star (red).}
\label{fig:hvarvar_radio}
\end{figure*}

Figure \ref{fig:hvarvar_radio} shows the effective net yields of C-12, N-14, O-16, Ne-22,\footnote{Ne-22 is an s-process neutron source massive stars and a neutron poison in AGB stars} Ni-56, and Fe-60 for various binary fractions, with assumed accretion efficiencies of $\beta_{\rm winds}=0.1$, $\beta_{\rm RLOF}=1$, and $\beta_{\rm SN}=0$. The round circles joined by lines are computed using CMCL4, while the non-joined red stars are computed using CMCL1. The reported single and primary yields from \cite{farmer2023} are shown in orange and red, respectively.

For C-12, N-14, O-16, and Ne-22, changing the binary fraction results in significant variation in the effective stellar yields. This is especially evident when when comparing the single ($h=0$) stellar yields with those of high binary fraction, and is increasingly significant as the primary mass increases. The large variation in O is particularly striking given that massive stars are expected to produce almost all O (whereas AGB stars produce most of the C, for example) \citep{iliadis2015}.

Ni-56 and Fe-60 also show significant variation in the net effective yield with the binary fraction, with the degree of variation also dependent on the primary mass. The binary-fraction dependence is of comparable significance to the mass-dependence. For most of the primary masses considered, the effective stellar yield tends to increase with increasing binary fraction, but there are regions of primary mass where this observation does not hold (these regions are different between Ni-56 and Fe-60). This reflects the somewhat chaotic nature of the single and primary supernova yields reported by \cite{farmer2023}, where there is often no simple description for whether the single or primary supernova yield for a given isotope is larger.

For many isotopes, the adopted model for the treatment of accreted material often affects the dependence on the binary fraction. For C-12, not conserving the primary's C-12 contributions (CMCL1) approximately halves the dependence on binary fraction. For O-16, it eliminates the binary fraction dependence almost completely. The binary fraction dependence also varies with the primary mass. N-14, for example, exhibits increasing effective net yields with binary fraction at lower initial primary mass but decreasing values at higher values.

\subsubsection{Wind accretion efficiency}

It is worth noting the effect of varying the different accretion efficiencies, shown in the lower panel of Fig. \ref{fig:mass_loss}. This figure presents the reported mass loss from each channel as a function of primary mass, as reported in \cite{farmer2023}, and provides highly relevant context to our discussion surrounding the effect of varying the accretion efficiencies.

The relative importance of each mass loss channel in terms of the raw amount of mass lost varies significantly with the primary mass. Below approximately 25 M\solar, most of the mass loss is through RLOF, with supernovae and winds making up only minority contributions. Beyond 30 M\solar, however, the situation is reversed, with most of the mass loss occurring through stellar winds (although both RLOF and the supernova continue to be significant sources of mass loss). It is expected that this behaviour will propagate to the effective yield calculations through the secondary yields. For example, increasing $\beta_{\rm winds}$ will most affect the effective stellar yields for the most massive stars, while the importance of $\beta_{\rm RLOF}$ will be higher for lower-mass stars. As a final note, we repeat the caution that the supernova mass loss is likely somewhat overestimated due to simplified explosion physics adopted in \cite{farmer2023}.

\begin{figure*}
\centering
\begin{subfigure}{0.99\columnwidth}
\centering
\includegraphics[width=0.99\columnwidth]{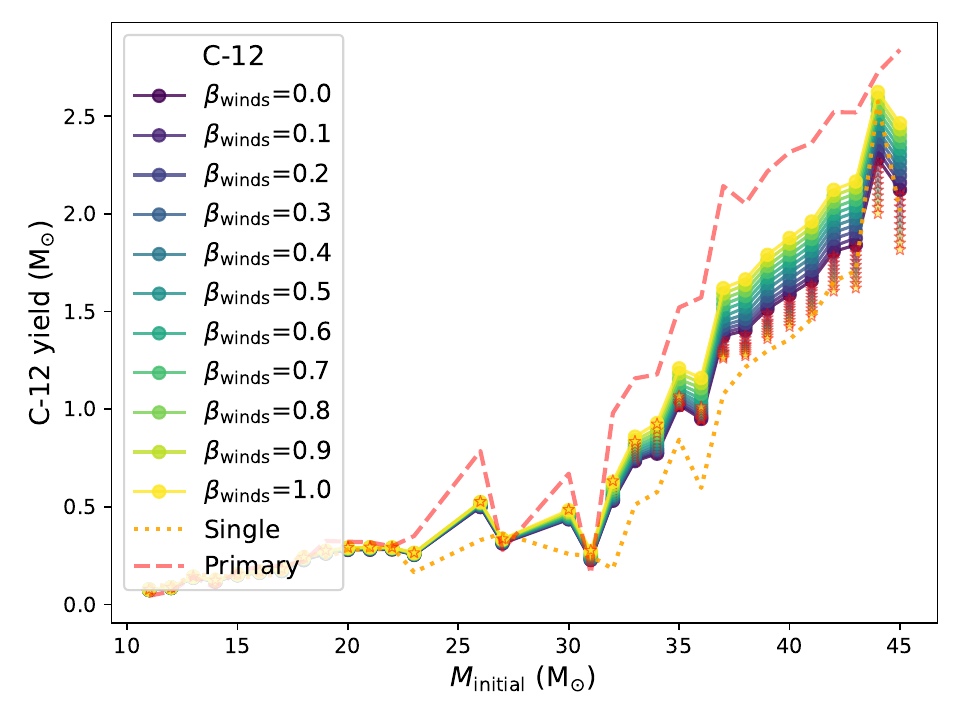}
\end{subfigure}%
\begin{subfigure}{0.99\columnwidth}
\centering
\includegraphics[width=0.99\columnwidth]{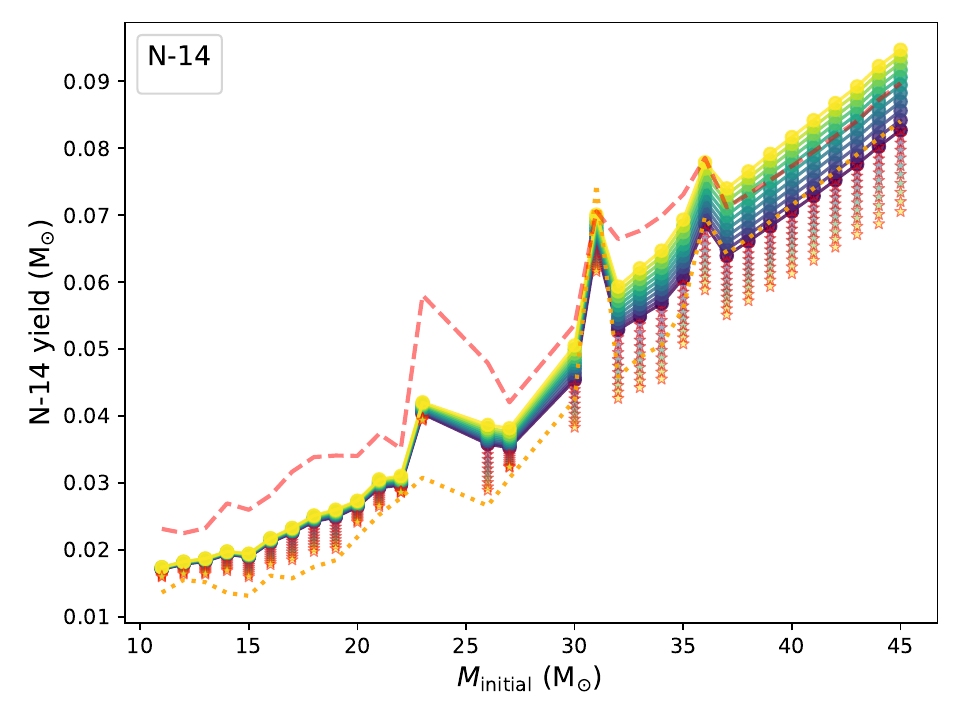}
\end{subfigure}

\begin{subfigure}{0.99\columnwidth}
\centering
\includegraphics[width=0.99\columnwidth]{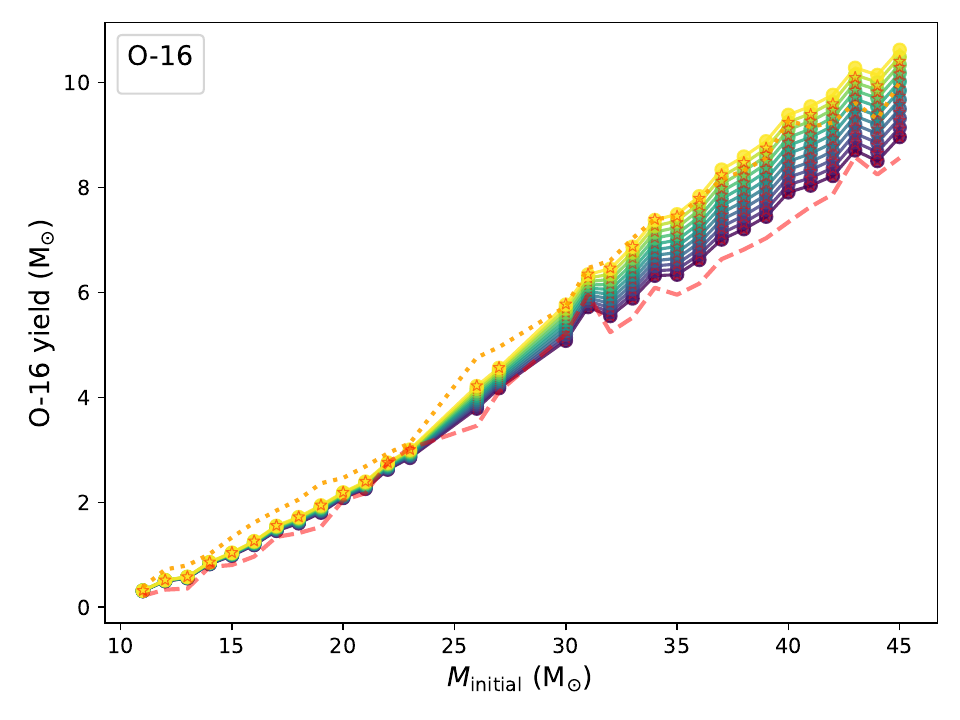}
\end{subfigure}%
\begin{subfigure}{0.99\columnwidth}
\centering
\includegraphics[width=0.99\columnwidth]{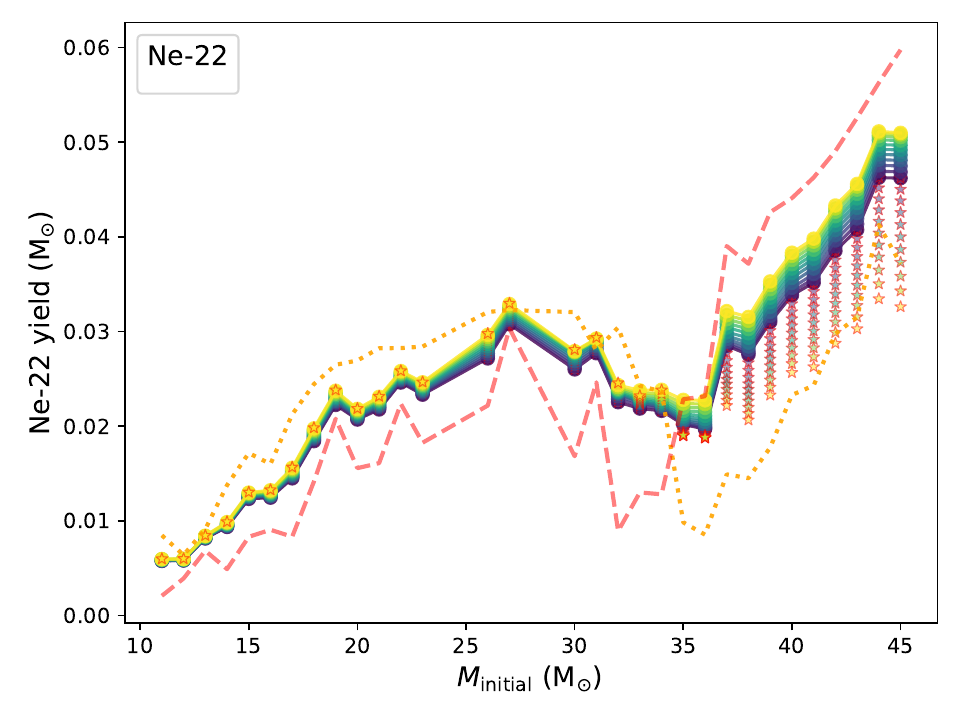}
\end{subfigure}

\begin{subfigure}{0.99\columnwidth}
\centering
\includegraphics[width=0.99\columnwidth]{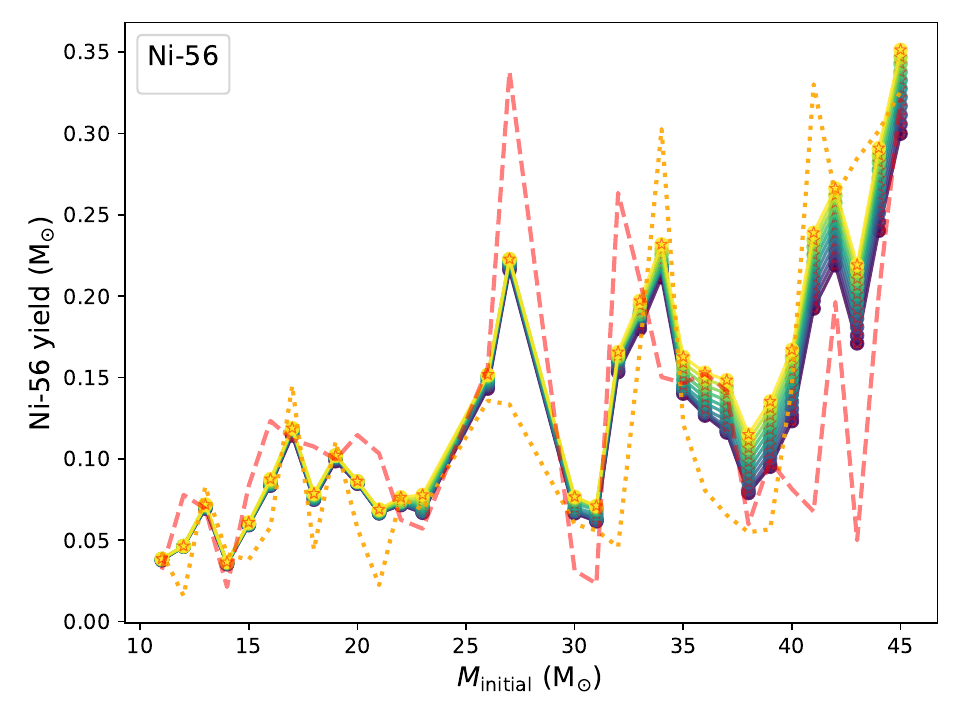}
\end{subfigure}%
\begin{subfigure}{0.99\columnwidth}
\centering
\includegraphics[width=0.99\columnwidth]{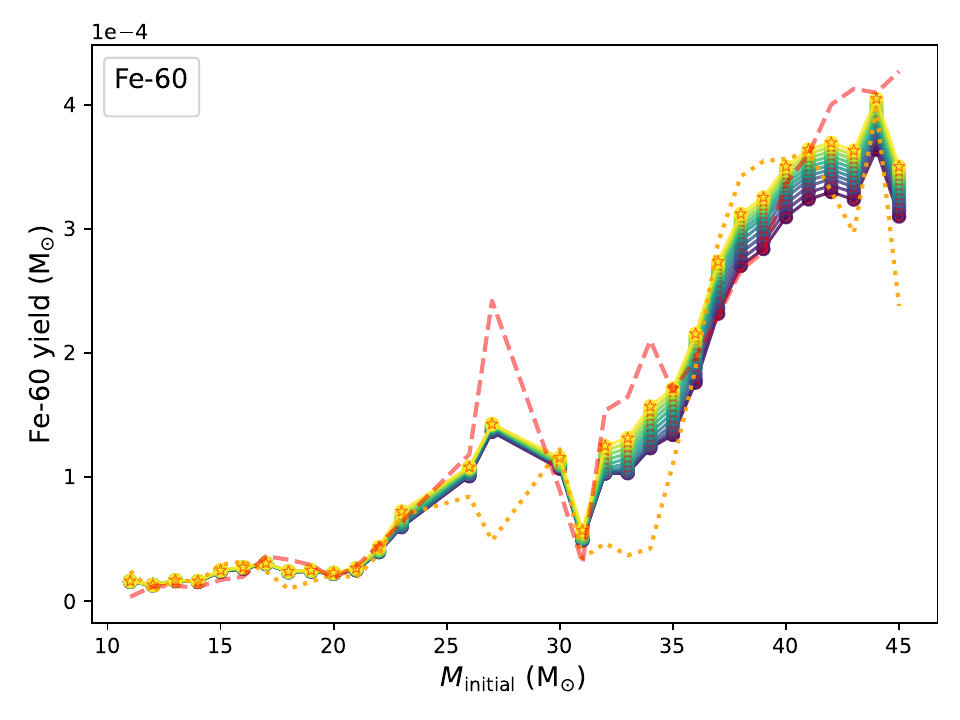}
\end{subfigure}

\caption{Effective net yields for C-12, N-14, O-16, Ne-22, Ni-56, and Fe-60, varying the wind accretion efficiency $\beta_{\rm winds}$ ($h=0.5$, $\beta_{\rm RLOF}=0$, and $\beta_{\rm SN}=0$). The connected circles were computed using CMCL4, while the isolated stars were computed using CMCL1. The dashed lines present the net yields of each isotope for the reported single star (orange) and the reported primary star (red).}
\label{fig:betawindvar_radio}
\end{figure*}

Fig. \ref{fig:betawindvar_radio} shows the effect of varying $\beta_{\rm winds}$ winds, assuming a binary fraction of $h=0.5$ and setting all other accretion efficiencies to zero. When increasing $\beta_{\rm winds}$, C-12, N-14, O-16, and Ne-22 all show noticeable increases in the effective yield, although when considering a more reasonable range of 0-20\% for $\beta_{\rm winds}$, this effect is small relative to the effect of changing the primary mass or binary fraction. The inverted (or reduced) relationship with $\beta_{\rm winds}$ in CMCL1 is reflective of the presence of processed C-12 and N-14 in the accreted material; by not conserving this production, the effective net yields decrease as more and more is artificially destroyed through accretion to the secondary.

The supernova-driven isotopes Ni-56 and Fe-60 both show noticeable variation beyond 30 M\solar, although this effect is secondary to the primary mass dependence even before considering that wind accretion efficiencies are unlikely to be greater than 20\%.

 \subsubsection{RLOF accretion efficiency}

The effect of varying $\beta_{\rm RLOF}$ is similar to that of varying $\beta_{\rm winds}$, although the range of reasonable values of $\beta_{\rm RLOF}$ differs significantly. Excluding systems experiencing extreme wind loss, we would expect accretion efficiencies upwards of 75\%. Furthermore, as shown in Fig. \ref{fig:mass_loss}, RLOF dominates the mass loss profiles for the primaries between 10 and 25 M\solar.

\begin{figure*}
\centering
\begin{subfigure}{0.99\columnwidth}
\centering
\includegraphics[width=0.99\columnwidth]{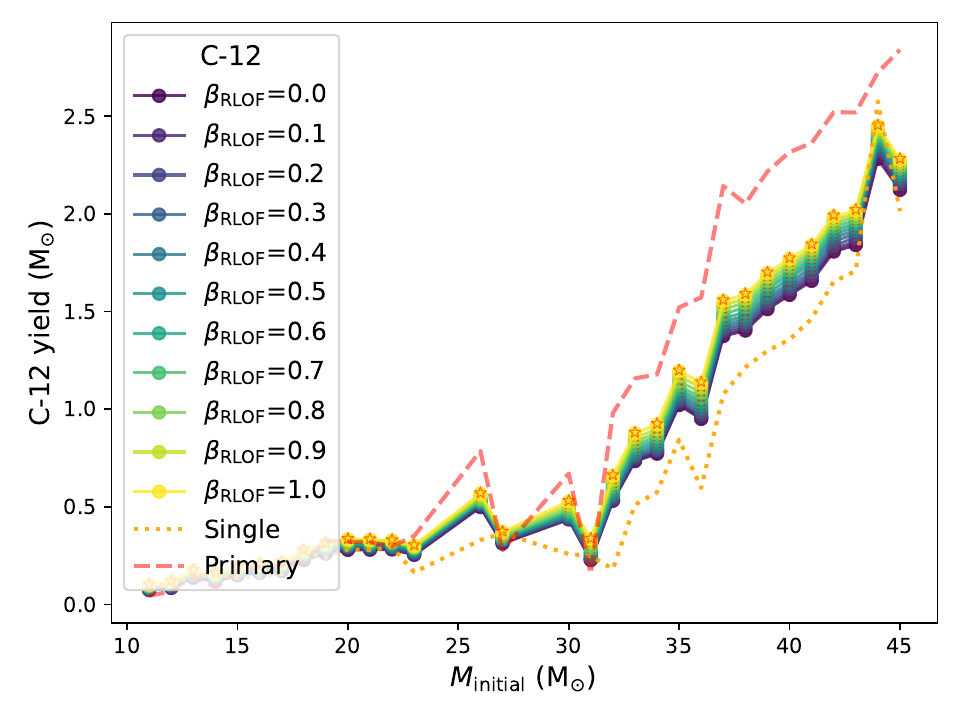}
\end{subfigure}%
\begin{subfigure}{0.99\columnwidth}
\centering
\includegraphics[width=0.99\columnwidth]{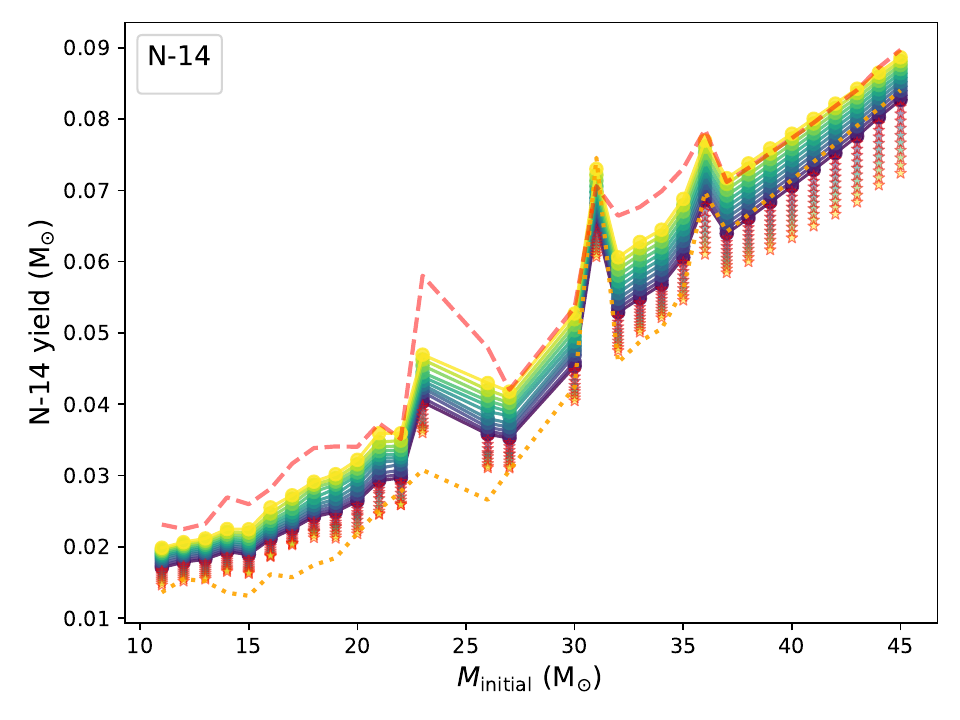}
\end{subfigure}

\begin{subfigure}{0.99\columnwidth}
\centering
\includegraphics[width=0.99\columnwidth]{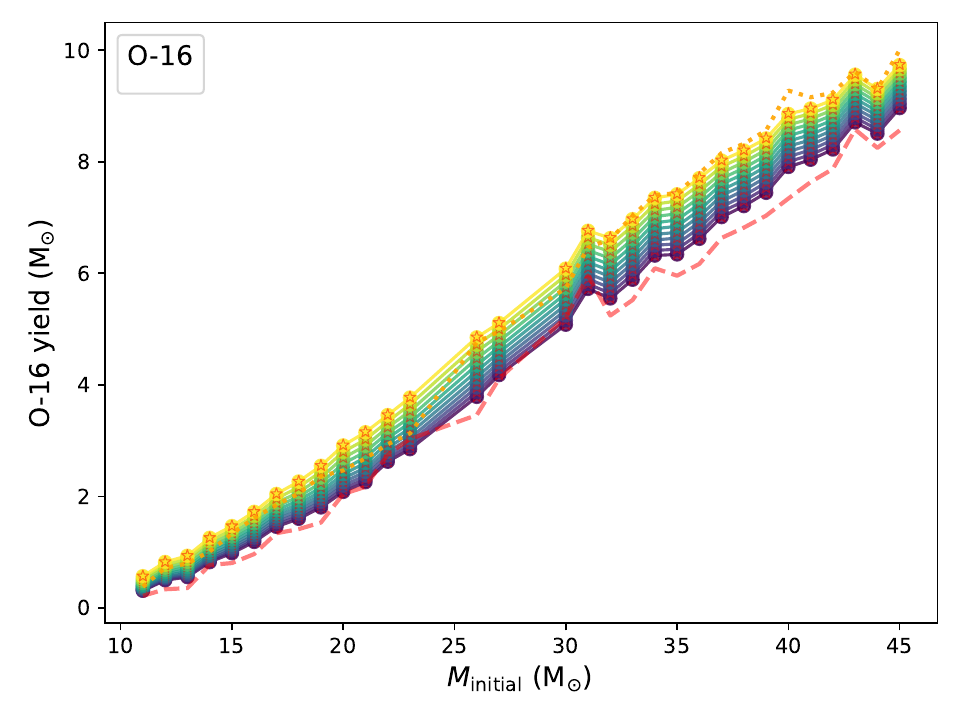}
\end{subfigure}%
\begin{subfigure}{0.99\columnwidth}
\centering
\includegraphics[width=0.99\columnwidth]{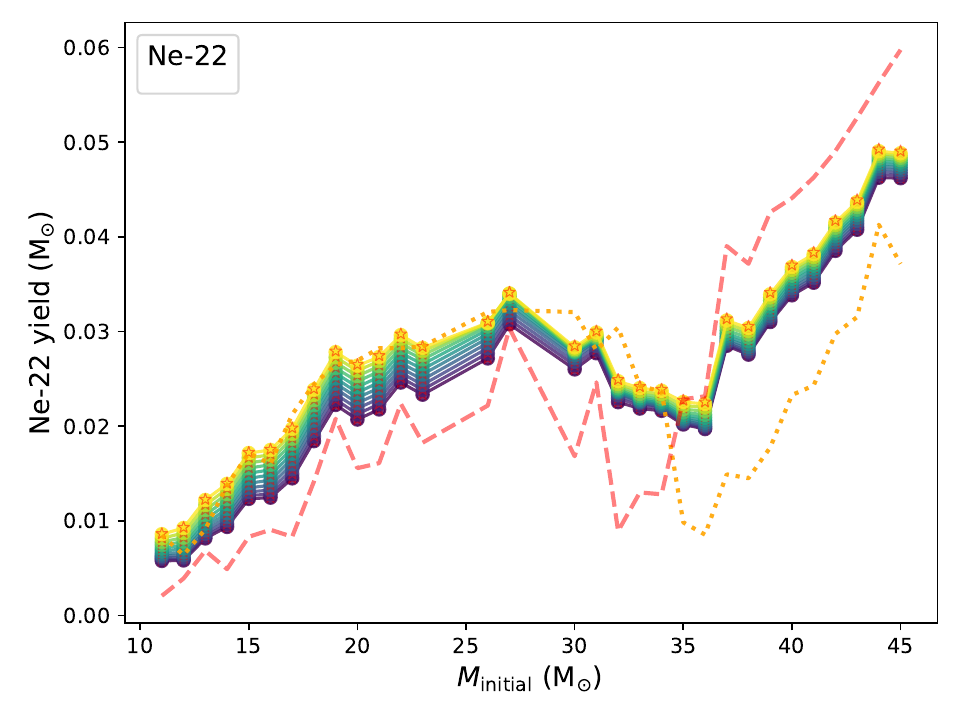}
\end{subfigure}

\begin{subfigure}{0.99\columnwidth}
\centering
\includegraphics[width=0.99\columnwidth]{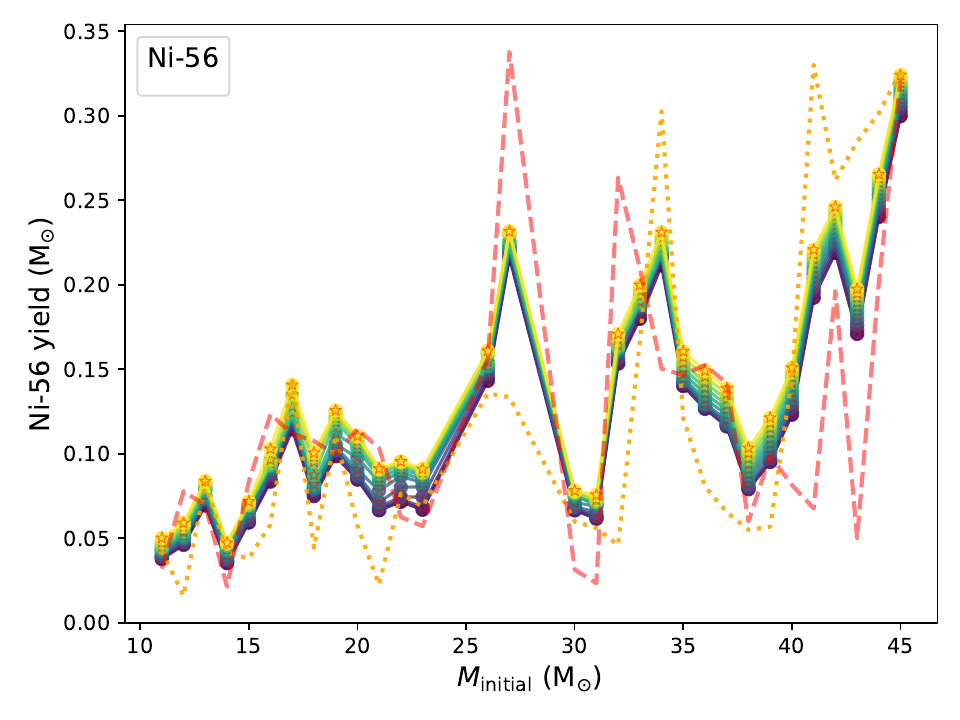}
\end{subfigure}%
\begin{subfigure}{0.99\columnwidth}
\centering
\includegraphics[width=0.99\columnwidth]{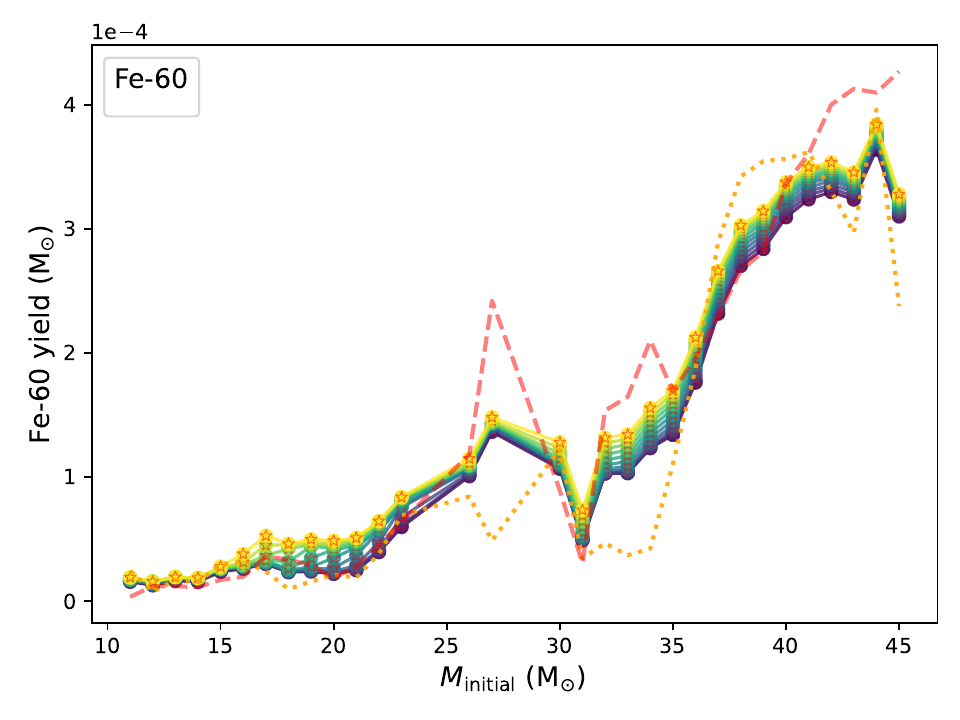}
\end{subfigure}

\caption{Effective net yields for C-12, N-14, O-16, Ne-22, Ni-56, and Fe-60 varying the RLOF accretion efficiency $\beta_{\rm RLOF}$ ($h=0.5$, $\beta_{\rm winds}=0$, and $\beta_{\rm SN}=0$). The connected circles were computed using CMCL4, while the isolated stars were computed using CMCL1. The dashed lines present the net yields of each isotope for the reported single star (orange) and the reported primary star (red).}
\label{fig:betarlofvar_radio}
\end{figure*}

With these considerations in mind, we find that the results shown in Fig. \ref{fig:betarlofvar_radio} are more or less as expected. We see increased variation in the effective yields of C-12, N-14, O-16, and Ne-22 with $\beta_{\rm RLOF}$ below 25 M\solar compared to the $\beta_{\rm winds}$ case, but somewhat less variation beyond 30 M\solar. This is also true for N-56 and Fe-60. 

\subsubsection{Supernova accretion efficiency}

Due to high ejecta velocities, supernova accretion efficiencies $\beta_{\rm SN}$ are extremely small. Indeed, it is likely that in a close binary material will be ablated from the companion. We therefore expect any amount of material accreted from the supernova to be negligible compared to the companion mass. However, any material accreted from such an event may be significantly enriched in certain isotopes.

\begin{figure*}
\centering
\begin{subfigure}{0.99\columnwidth}
\centering
\includegraphics[width=0.99\columnwidth]{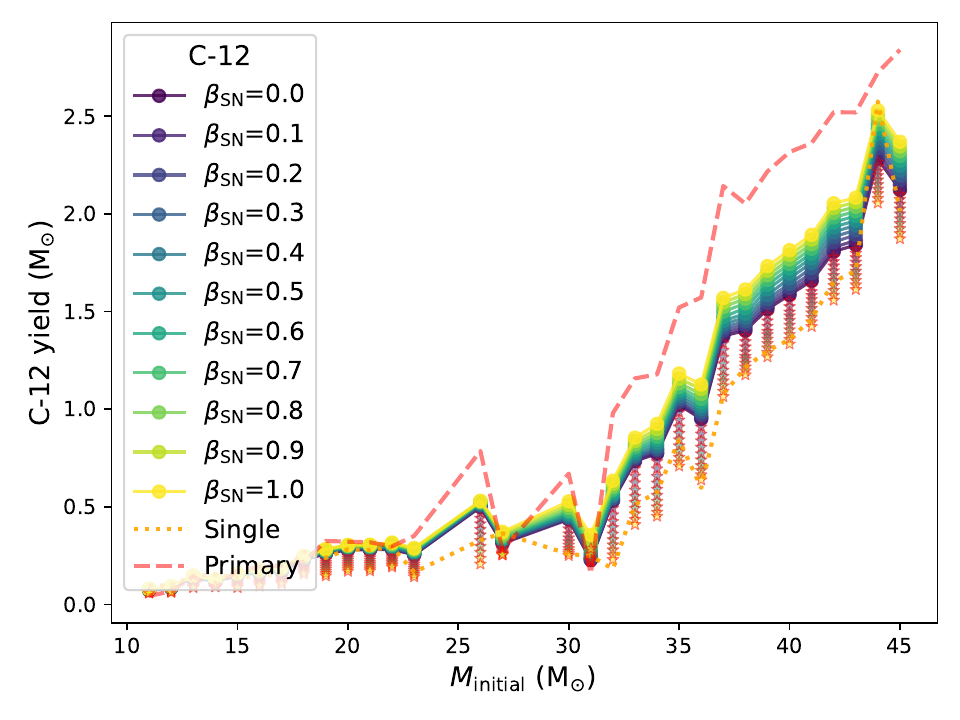}
\end{subfigure}%
\begin{subfigure}{0.99\columnwidth}
\centering
\includegraphics[width=0.99\columnwidth]{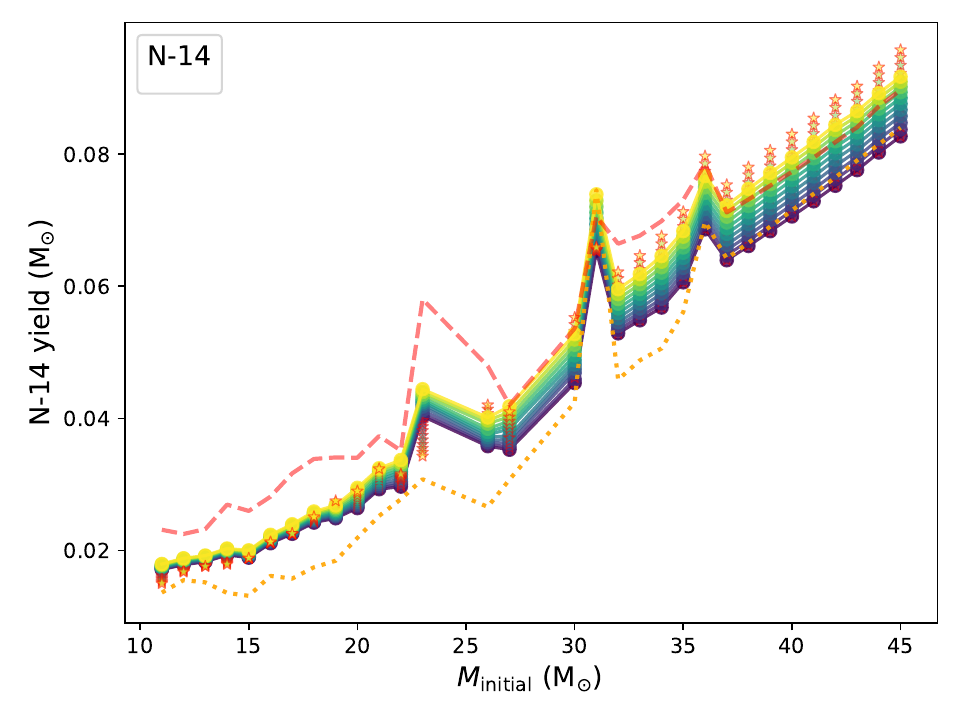}
\end{subfigure}

\begin{subfigure}{0.99\columnwidth}
\centering
\includegraphics[width=0.99\columnwidth]{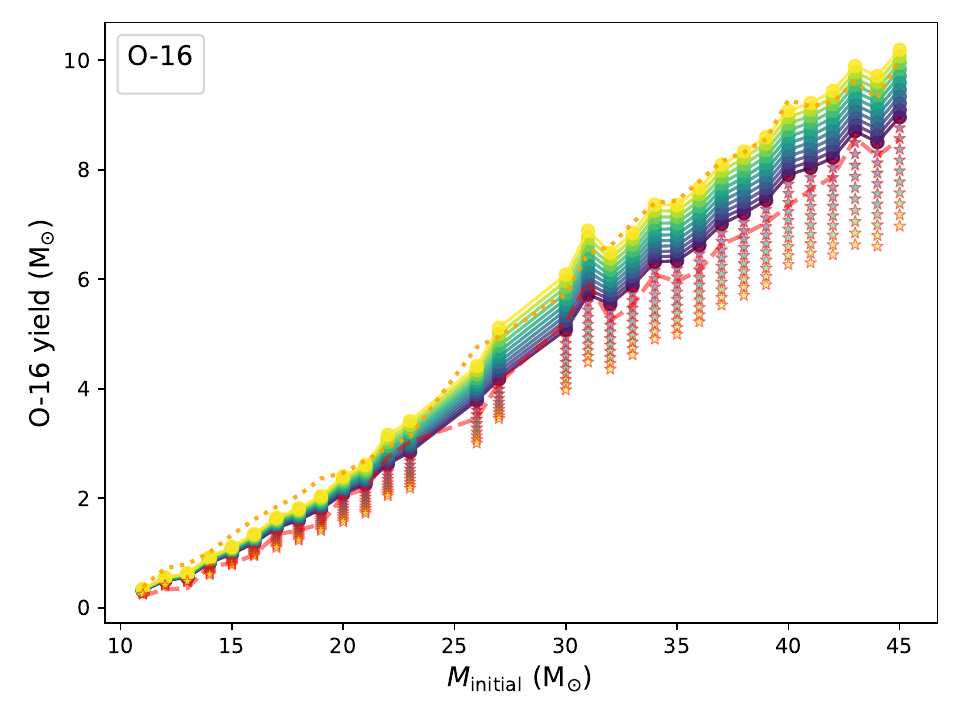}
\end{subfigure}%
\begin{subfigure}{0.99\columnwidth}
\centering
\includegraphics[width=0.99\columnwidth]{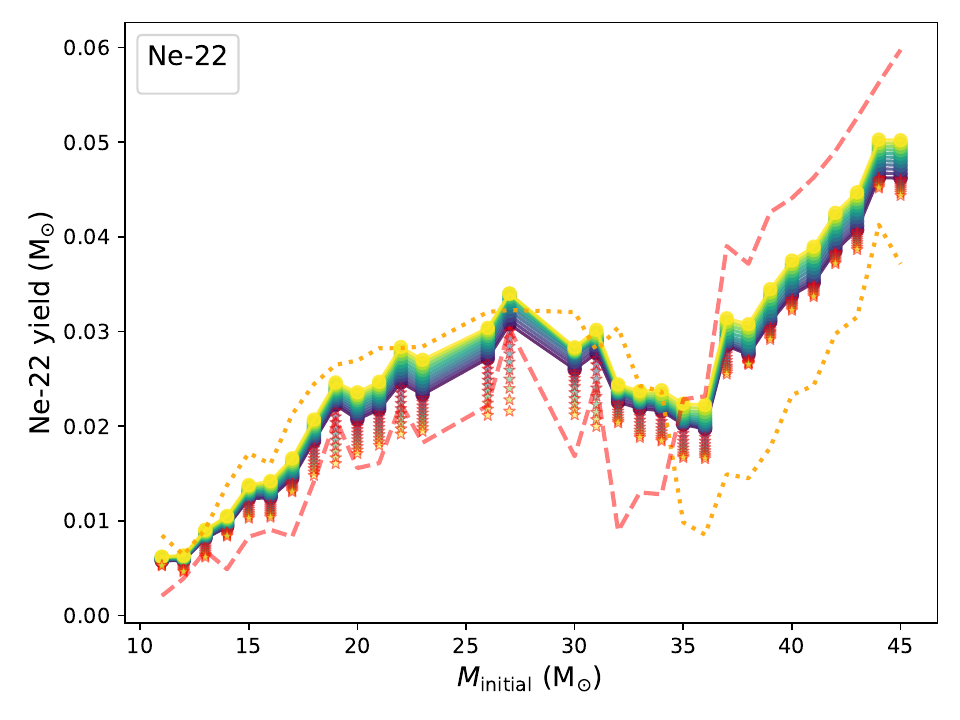}
\end{subfigure}

\begin{subfigure}{0.99\columnwidth}
\centering
\includegraphics[width=0.99\columnwidth]{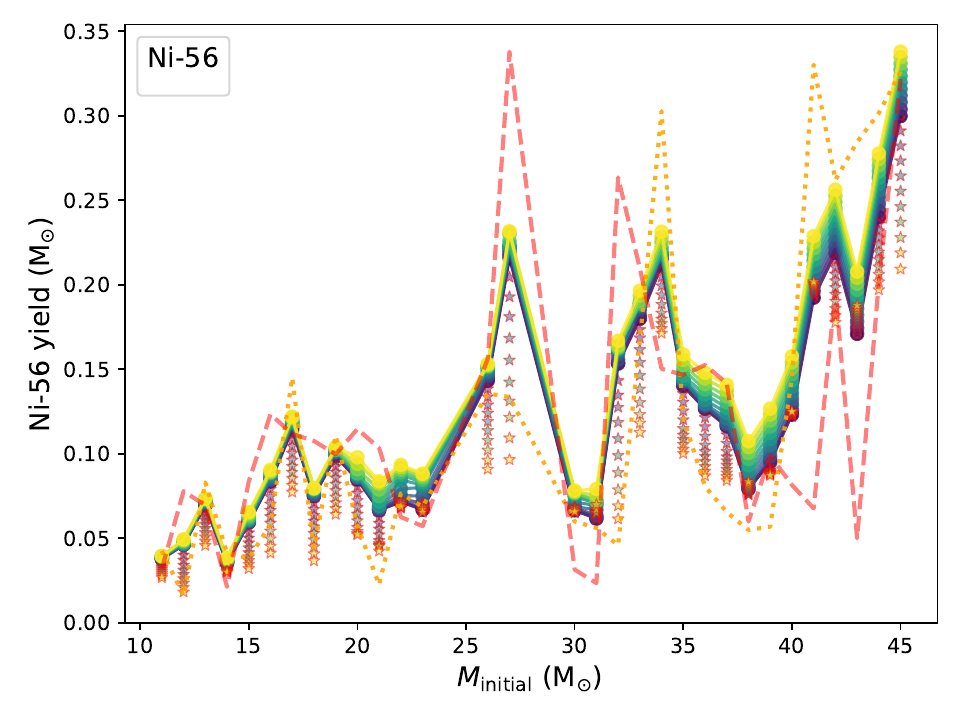}
\end{subfigure}%
\begin{subfigure}{0.99\columnwidth}
\centering
\includegraphics[width=0.99\columnwidth]{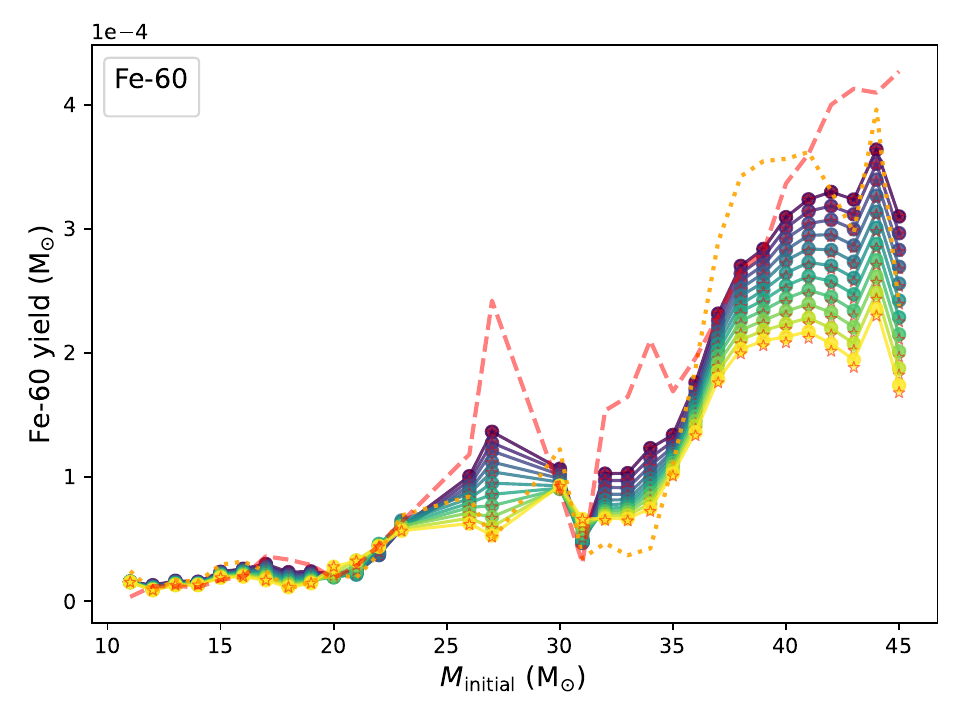}
\end{subfigure}

\caption{Effective net yields for C-12, N-14, O-16, Ne-22, Ni-56, and Fe-60 varying the supernova accretion efficiency $\beta_{\rm SN}$ ($h=0.5$, $\beta_{\rm winds}=0$, and $\beta_{\rm RLOF}=0$). The connected circles were computed using CMCL4, while the isolated stars were  computed using CMCL1. The dashed lines present the net yields of each isotope for the reported single star (orange) and the reported primary star (red).}
\label{fig:betaccvar_radio}
\end{figure*}

Examining Fig. \ref{fig:betaccvar_radio}, we see that for C-12, N-14, O-16, and Ne-22 the level of variation is similar to that of varying $\beta_{\rm winds}$ or $\beta_{\rm RLOF}$. Under the assumptions of CMCL4, the effect of adjusting the supernova efficiency for most isotopes can be understood in terms of the altered evolutionary mass of the secondary.

Ni-56 shows relatively little variation in CMCL4 but significant variation in CMCL1, as we would expect given its production during the supernova. Most of the $\beta_{\rm SN}$-dependent variation  that is present (in CMCL4) occurs at the highest masses. In CMCL1, Ni-56 shows significant variation at all masses due to its relatively flat mass dependence. Fe-60, on the other hand, shows a strong mass dependence and high levels of variation with $\beta_{\rm SN}$. As previously discussed, CMCL4 and CMCL1 are in lock-step for the short-lived radioisotopes when varying accretion efficiencies to CMCL4 accounting for their decay during the remaining lifetime of the secondary. This decay results in the strong decline in Fe-60 production with increasing $\beta_{\rm SN}$, as the primary's contribution to the ISM is increasingly captured by the secondary where primary's Fe-60 has sufficient time to almost completely decay before its eventual expulsion into the ISM by the secondary. For none of the isotopes considered do we see evidence of significant variation in the net isotopic yields under the assumptions of CMCL4 when considering the narrow range of physically reasonable supernova accretion efficiencies. 

The lack of nuclear-post-processed yields for the binary stellar models places r-process nucleosynthesis beyond the scope of this work. We note, however, that it seems unlikely that whether the true supernova accretion efficiency is 0 or 0.1 will significantly affect the effective stellar yield calculation even for these isotopes. Rather, the effect of binarity will express itself through the blending of the primary, secondary, and single star yields in the effective stellar yield calculation. The statistical distribution this forms will naturally depend on each star's pre-supernova configurations, which will be sensitive to all accretion efficiency parameters and the binary fraction, but without any particular sensitivity to $\beta_{\rm SN}$.

\subsubsection{Elemental abundance ratios}

\begin{figure*}
\centering

\begin{subfigure}{\textwidth}
    \centering
    \includegraphics[width=\textwidth]{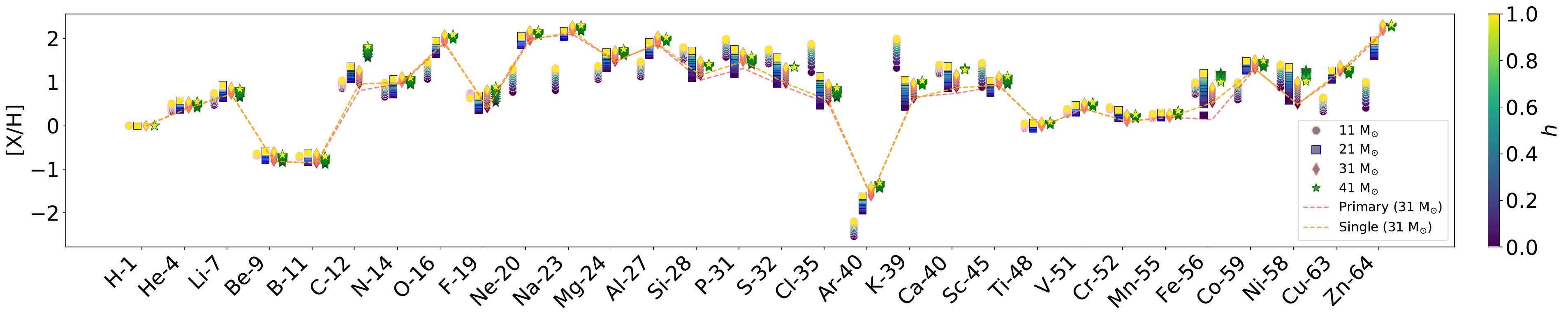}
\end{subfigure}

\begin{subfigure}{\textwidth}
    \centering
    \includegraphics[width=\textwidth]{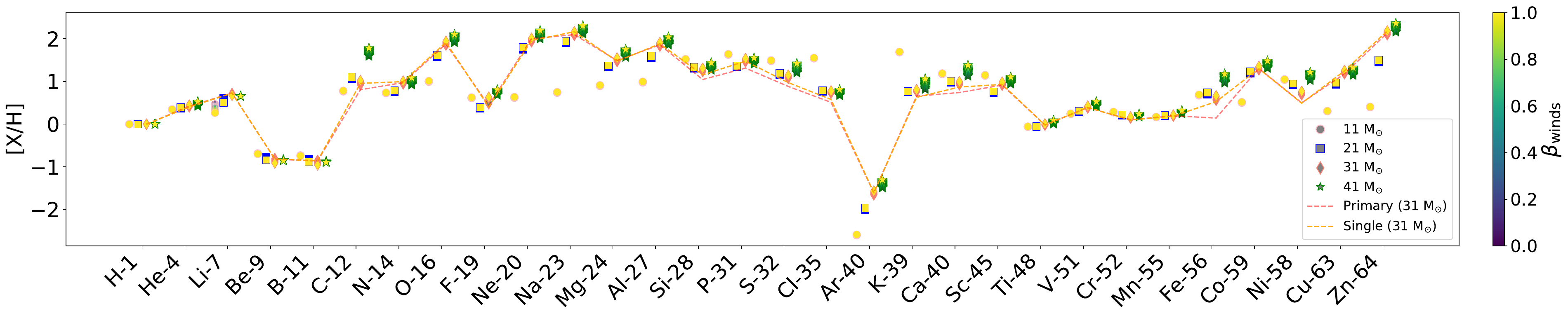}
\end{subfigure}

\begin{subfigure}{\textwidth}
    \centering
    \includegraphics[width=\textwidth]{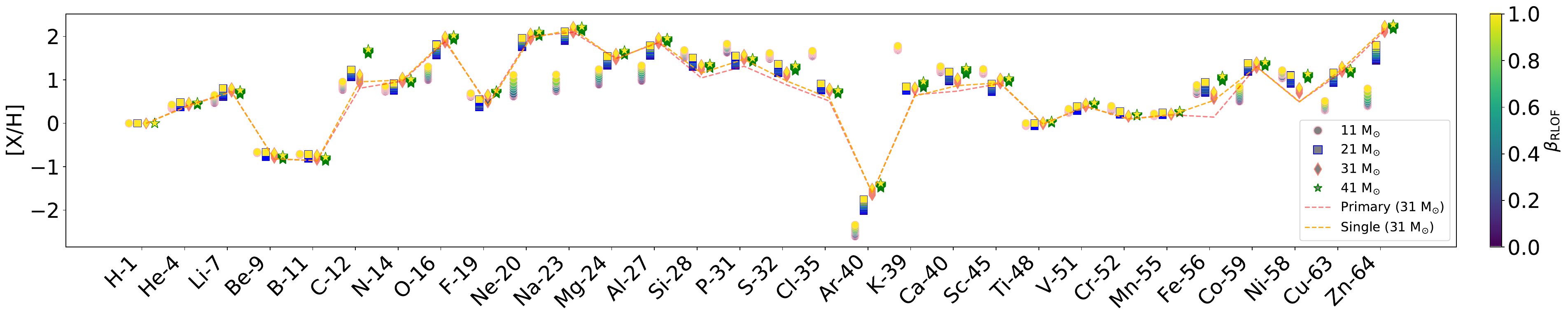}
\end{subfigure}

\begin{subfigure}{\textwidth}
    \centering
    \includegraphics[width=\textwidth]{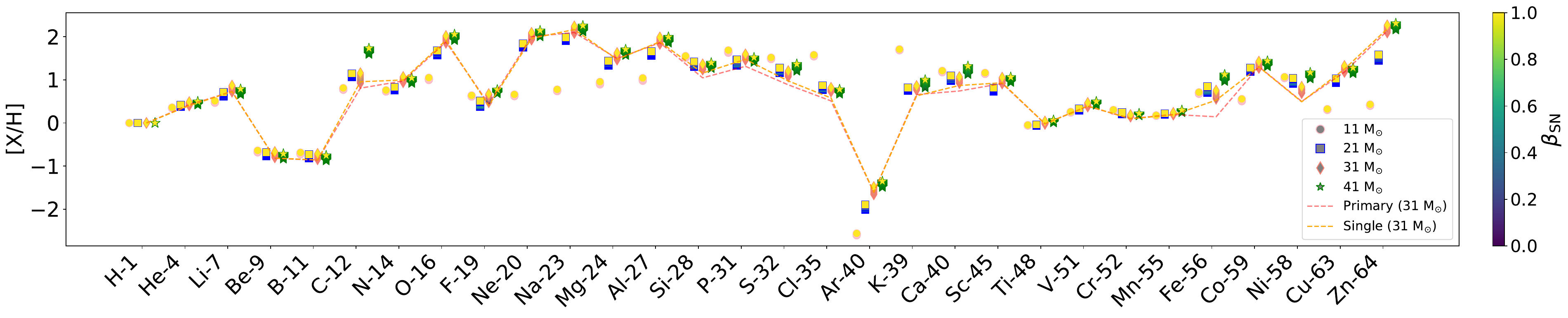}
\end{subfigure}

\caption{[$X$/H] calculated for 11~M\solar, 21~M\solar, 31~M\solar, and 41~M\solar\ effective stellar yields (coloured markers). From top to bottom, the panels show the effect of varying $h$, \sbs{\beta}{winds}, \sbs{\beta}{RLOF}, and \sbs{\beta}{SN}. [$X$/H] calculated for a 31 M\solar\ star from \cite{farmer2023}'s single and binary models is also plotted for comparison.}
\label{fig:X_H_subfigures}
\end{figure*}

Figure \ref{fig:X_H_subfigures} shows logarithmic isotopic number abundance ratios relative to the solar values (${[X/{\rm H}]=\log(N_X/N_{\rm H})_* - \log(N_X/N_{\rm H})}$\solar) for the dominant stable isotopes\footnote{In Fig. \ref{fig:X_H_subfigures} we use the reported Ni-56 yields for Fe-56} of all elements with yields calculated in \cite{farmer2023}. These ratios can be taken as close approximations to the elemental abundance ratios of most elements. We show [$X$/H] calculated using effective stellar yields for 11~M\solar, 21~M\solar, 31~M\solar, and 41~M\solar\ effective stellar yields for different binary fraction and accretion efficiency parameters. The values of [$X$/H] calculated for a 31 M\solar\ star from \cite{farmer2023}'s single and binary models directly are also shown for comparison.

We find that there can be a significant impact on the [$X$/H] abundance ratios when varying the binary fraction $h$. Higher binary fractions act to drive up the abundance ratios relative to H of most elements. The binary fraction-dependent variation is most significant for lower-mass primaries, where we see it varying by as much as 0.5-1~dex for the 11~M\solar\ and 21 M\solar\ examples shown in Fig. \ref{fig:X_H_subfigures}. For some isotopes, this level of variability exceeds the primary-mass dependent variation (e.g. Si-28 and P-31). The higher-mass primaries exhibit significantly lower levels of variation, typically less than 0.3~dex for 31 M\solar, and <0.1~dex for 41 M\solar.

Compared to varying $h$, varying the accretion efficiency generally results in far less of an effect on [$X$/H]. This lowered impact on the abundance ratios is reflective of the lower variation in the yield calculations. The effect of increasing the accretion efficiency is almost always to drive up the [$X$/H] abundance ratios, albeit often only very slightly.

When varying $\beta_{\rm winds}$, the level of variation negligible for all but the 41 M\solar\ example, as we would expect from inspection of Fig. \ref{fig:mass_loss}. Even for the 41 M\solar\ case, the level of variation is typically less than 0.2~dex. Varying $\beta_{\rm RLOF}$ results in the greatest level of $\beta$-dependent variability. This dependence is greatest for the 11~M\solar\ and 21~M\solar\ examples, where it can reach around 0.5~dex for some isotopes. The level of variation for the higher-mass stars is lower, less than 0.2~dex. Varying $\beta_{\rm SN}$ results in similar behaviour to varying $\beta_{\rm winds}$, with more massive stars exhibiting higher variation, but still at a low level ($<0.2$~dex).

\section{A checklist for binary stellar modellers}
\label{sec:checklist}

In Section \ref{sec:effective_yields}, we discussed how detailed yields from the binary stellar evolution calculations can be included into GCE. However, our ability to take full advantage of what these detailed calculations can provide is limited by the kinds of output that is reported. Further, the completeness of such a mixed-population GCE model will be limited by the mass, metallicity, and evolution channels for which yields have been computed, as well as inheriting all of the modelling uncertainty from the underlying stellar models used to compute those yields. In this section, we first provide a checklist for modellers reporting outputs for binary stellar channels, before outlining where additional additional computations for different evolution channels are most needed.

\subsection{Output checklist for binary stellar modellers}
\label{sec:checklist_output}
We  first provide the full checklist for convenience, and then discuss each item subsequently.

\begin{enumerate}
    \item [I.] Report the amount of each isotope lost through winds, RLOF, and supernovae.

    \item [II.] Calculate and report supernova yields using the final state of the binary evolution models. 

    \item[III.] Provide an equivalent set of single stellar yields computed using comparable physical and numeric assumptions.

    \item[IV.] Provide representative timings for the different mass loss episodes.

    \item[V.] Publish the data in a machine-readable format through publicly accessible archives.
\end{enumerate}

\subsubsection*{I. Report the mass lost through different mass loss mechanisms}

Much of the complexity -- and opportunity for investigating binary stellar evolution effects -- in our treatment for stable mass transfer onto a main sequence companion is introduced in our estimation of the yields of the secondary star. We were able to account for variable binary physics because of the complete way that \cite{farmer2023} report their data. Instead of simply providing the total yield of each isotope for each primary mass, they provide a breakdown of the yields by mass-loss channel (through winds, RLOF, or supernova explosion). Their method of doing so, providing both the total mass of each isotope lost through each channel and the mass of that isotope that was already present in the star at birth, can be considered ideal. This information allows different values for the conservativeness of mass transfer for these different channels to be considered, as well as the development of more sophisticated methods for treating the deposition of radioactive material onto the accretor.

At minimum, the amount of mass lost through RLOF is required to approximate the correct average secondary yield. Without this minimum requirement, the term for the average secondary yield will be considerably in error.

\subsubsection*{II. Calculate and report supernova yields}

The calculation of bespoke supernova yields is essential for binary massive stars. Binary evolution will significantly affect the conditions under which the resulting supernova of the primary occurs, as it alters -- at minimum -- the amount of envelope left on the star at the time of supernova, as well as the compactness factor of the core, a key property for the evolution of the supernova. These supernovae are fundamentally different from those of single star models, making attempts to `graft' other supernova models onto these binary stellar calculations poorly motivated and unlikely to capture the true impact of binary stellar evolution on the supernova yields. Given the dominance of supernovae (compared to massive stellar winds) for many elemental yields, such a computation can be expected to miss much of the impact of binary stellar evolution on elemental abundances.

Thus far, binary stellar evolutionary models with stellar yields have been calculated using 1D stellar evolution calculations using codes such as \texttt{MESA} \citep{Paxton2011,Paxton2013,Paxton2015,Paxton2018,Paxton2019}. These codes allow the consistent calculation of the supernova progenitor, but are not ideal for computing the inherently multi-dimensional supernova itself. Ideally, binary progenitor models would be used to initialise 3D supernova hydrodynamic simulations to compute the binary supernova yields. Practically speaking, this demands a different skill-set to running binary stellar evolution calculations, and also represents a significant increase in the computational cost of computing a grid of binary stellar yields \citep{boccioli2024review}.

{\cite{farmer2023} calculate 1D explosions in MESA to estimate the supernova yields, a strategy that is likely the most practical solution for future binary stellar modellers. They do, however, make one strong simplifying assumption when doing so that compromises the fidelity of their results: they set a constant supernova energy for all progenitor models. This assumption is unrealistic; in nature, the supernova energy should depend on the progenitor's structure. \cite{farmer2023} acknowledge the effect this assumption has on their (unusually low) remnant masses, with the high explosion energies reducing fall-back to negligible quantities \citep{goldberg2019}.

In the effective stellar yield framework, the issue of explodability can be dealt with by excluding the supernova term from the calculation of the effective yield for certain progenitors. The explodability of different mass regimes of massive stars remains highly debated in stellar modelling circles \citep{sukhbold2016,boccioli2024}. Although it has been suggested that binary stripped stars may explode more easily due to lower core compactness \citep{vartanyan2021}, binary progenitors remain understudied. We do not enforce an explodability model in this work, in order to permit the user to make the decision of which explodability condition, if any, to enforce upon the primary and secondary. Ideally, binary-stripped star models should motivate the explodability of the primary, while the secondary can likely be best treated by an explodability model derived from single star models.

\subsubsection*{III. Provide a complementary set of single stellar yields.}

In order to isolate the binary evolution from effects from physical -- or numeric -- differences between different sets of stellar models, ideally the yields for the single stars and the binary stars should be computed according to the same physical and numeric methodology. The provision of a range of single stellar models incorporating different rotation rates would be even better, allowing not only a more complete definition of single stellar yields, but also resulting in more accurate yields for the secondary for many non-merger binary evolution channels.

\subsubsection*{IV. Provide representative timings for mass loss through each channel.}

\textbf{For astrophysical problems such as the evolution of the early Solar System, small differences in the timing of the yields may become important.} In Section \ref{sec:effective_yields} we assumed that that all mass loss occurred at the end of the star's life. Wind mass loss often occurs occur over a long time period, but there may also be a significant offset between the mass transfer phase and the supernova in some cases, and when considering more complex evolutionary channels involving stellar mergers and/or common envelope events, the timings will become more complicated. For this reason we strongly encourage modellers to report, for each model, the timing of the onset of significant mass loss through each channel as well as the time of peak mass loss through that channel.

\subsubsection*{V. Machine readable tables and organised, publicly accessible data-sets.}

Finally, we end this list not on a scientific recommendation but a practical one. It is essential that the diverse information needed to build sophisticated GCE models is freely available and accessible. Publishing scientific data online in an accessible (machine-readable) format adds enormous additional value to that science, ensuring it will benefit the scientific community for years to come. We recommend the reporting structure of \cite{farmer2023} for those seeking guidance on the level of completeness and structure, and note that those authors also make available online many of the scripts required to output \texttt{MESA} output in such a format.

\subsection{Additional models needed for completeness}
\label{sec:checklist_models}

To fully capture the complexity of binary stellar evolution and its effect on the chemical evolution of real stellar environments, binary interactions including binary stripping, accretion, and stellar mergers should be accounted for across the traditional stellar mass and metallicity parameter space. To achieve this level of completeness, significant numbers of additional binary yield calculations are needed beyond what have been computed thus far. Following the structure of the previous section, we provide a summarised list before discussing each item in detail.

\begin{enumerate}
    \item [I.] Binary yields for low-mass stars.
    \item [II.] Metallicity-dependent binary yields.
    \item [III.] Mass gain: stable mass transfer and stellar mergers.
    \item [IV.] Mass donors
    \item [V.] Complex binary evolution
    
\end{enumerate}
 
\subsubsection*{I. Binary yields of low-mass stars.}

Despite AGB stars being long established as essential sites of C, N, and the slow neutron capture process (s-process) production (e.g. \citealt{karakas2016}), there is a dearth of detailed studies reporting yields from binary stellar evolution channels. Synthetic populations of low-mass binary stars with yield estimates have been produced on several occasions (e.g. \citealt{izzard2006,osborn2023}), but stellar yields from detailed binary evolutionary calculations similar to those undertaken by \cite{farmer2023} and \cite{brinkman2023} have, to the authors' knowledge, never been computed.\footnote{\cite{tout2000} consider the effect of binary stellar evolution on C production in AGB stars, concluding that avoided AGB phases will be the dominant effect.}

While binary population synthesis calculations can offer a more complete picture of the binary stellar pathways, their ability to accurately predict the effect of binary stellar evolution on the yields is limited and can be discrepant from detailed evolutionary calculations. \cite{osborn2023} note that according to their binary population synthesis calculations, stellar mergers between first-ascent giant branch stars and main sequence stars result in the post-merger star entering the AGB phase with an under-massive core relative to its envelope, ultimately resulting in significantly increased Al-26 (produced in hot-bottom burning, see \citealt{boothroyd1992,lattanzio1992,doherty2014,karakas2016}) yields due to an extended AGB phase. However, when verifying this result using detailed models, \cite{osborn2023} found that the memory of the merger on the giant branch was effectively erased during core He burning, and its subsequent AGB evolution mirrored that of a single star of equivalent post-merger mass.

For a variety of reasons, ranging from the sensitivity of s-process elements to the details of C-13
pockets and inter-shell mixing to the number of thermal pulses occur prior to mass loss, a dedicated grid of detailed binary evolutionary models including AGB nucleosynthesis is urgently needed. These models should cover binary-induced mass gain and loss at various evolutionary phases. A set of such models covering masses relevant to AGB stars would form an excellent starting point for accounting for binary stellar evolution in low-mass stars. 

\subsubsection*{II. Metallicity dependent binary yields}

Thus far, only solar metallicity ($Z\approx0.014$, \citealt{asplund2009}) binary stellar yields have been computed. There are no binary stellar yields computed at non-solar metallicities, a bias that is unrepresentative of the historical populations that have contributed to the chemical evolution of our galaxy and the wider universe. It has long been established that the metallicity significantly affects both stellar evolution and the stellar yields. Thus, the absence of low-metallicity binary stellar yields represents a significant hurdle to the proper use of binary stellar yields within GCE codes, forcing users to adopt measures such as scaling existing metallicity-dependent single stellar grids with binary-to-single ratios in order to explore the effect of binary stellar evolution on the chemical evolution of real stellar environments (see e.g. \citealt{storm2025}).

As the already considerable work load of computing a grid of binary stellar yields scales linearly with the metallicity resolution, being strategic with the metallicities computed is advantageous. Considering the need to intersect with binary population synthesis codes to determine the weightings for different evolution channels, a grid of metallicities that overlaps with the metallicities accessible to binary population synthesis simulations from 1.5~\Z\solar\ to 0.005~\Z\solar\ is needed. Such a grid would provide full coverage of the ubiquitous \texttt{SSE} fitting formulae \citep{pols1998,hurley2000,hurley2002} that most binary population synthesis codes rely upon.\footnote{The advent of interpolation-based synthetic stellar evolution codes such as SEVN \citep{spera2019}, METISSE \citep{agrawal2020}, and POSYDON \citep{fragos2023} may soon render this metallicity consideration obsolete.}

\subsubsection*{III. Mass gain: Stellar mergers and stable mass transfer}

One-dimensional stellar structure codes do not handle mass gain very well, particularly if the star is allowed to spin up. Being hydrostatic evolution codes, they also have little chance of accurately modelling hydrodynamic effects that may become important during stellar mergers. A combined 3D and 1D approach would seem intuitive, but these computations are so expensive that computing a suitable grid of 3D models is impractical even with a 1D code handling the pre- and post-merger evolution. 

Despite these challenges, efforts to compute the evolution of (non-rotating) post-merger stars in 1D are ongoing, demonstrating significant -- and even potentially asteroseismically observable -- structural differences between post-merger stars and single stars of similar mass \citep{henneco2024}. Unfortunately, the rapid-accretion method employed by \cite{henneco2024} and other such studies (e.g. \citealt{rui2021}) result in systematic under predictions for the amount of processed material due to the assumption of accretion of material matching the surface composition of the accretor. An alternative merger approximation such as the entropy-sorting method \citep{gaburov2008} is likely needed to accurately predict the stellar yield of the eventual post-merger product.

Technical challenges aside, the common-place nature of stellar mergers \citep{podsialowski1992,demink2014} makes them vital to forming a complete model of binary stellar evolution. The outcome of a stellar merger calculation will depend both on the masses and the evolutionary states of the two pre-merger stars. 

We suggest the following combinations should be prioritised: main sequence + main sequence; stars with non-degenerate He cores (e.g. Hertzsprung gap) + main sequence stars; stars with degenerate He cores (e.g. late first-ascent giant branch) + main sequence stars; core He burning stars + main sequence stars; stars which have completed He burning + main sequence stars. In each case, a range of stellar mass combinations should be considered, spanning the physically reasonable range for each scenario. This would provide an excellent starting point, covering many of the most relevant scenarios. Of secondary priority would be the more exotic mergers between evolved stars and stellar remnants (white dwarfs, neutron stars, and black holes) and of tertiary priority would be mergers between two evolved (non-remnant) stars.

Stable mass accretion can be dealt with far more cheaply. Simply due to where stars spend their lives, the vast majority of stable mass accretors will be either main sequence stars or stellar remnants. Stable mass transfer onto an evolved star requires an even larger evolved star to be co-existing, which in turn generally requires a very high value of the initial mass ratio. Material processed by stably accreting neutron stars and black holes probably only rarely makes its way back into the interstellar medium, and (unlike merging neutron stars, \citealt{kobayashi2023nsm}) are not currently considered viable sites for most nucleosynthesis. This channel could still manifest in an effective stellar yield calculation as a fraction of secondaries themselves undergoing binary stripping where the companion star acts purely as a sink rather than a delayed source of nuclear material. Stably accreting white dwarfs will output material processed through the hot-CNO cycle (see e.g. \citealt{starrfield1978,jose1998,iliadis2015}) in nova eruptions, which can best be included as extra terms in GCE calculations \citep{kemp2022li,kemp2024}, as opposed to being folded into an effective stellar yield calculation.

\subsubsection*{IV. Mass donors}

The effect of mass loss through stable mass transfer is investigated by \cite{farmer2023}, \cite{brinkman2023}, and \cite{nguyen2024}, although \cite{nguyen2024} do not report complete yield sets. \cite{farmer2023} exclusively look at mass loss from an evolved star, while \cite{brinkman2023} consider a range of different orbital periods, resulting in mass transfer occurring during different phases of evolution, which they subdivide into mass transfer occurring when the donor is either on the main sequence (case A) or after the main sequence, but before core He burning (case B). Although small variations in the yields within each mass transfer case do exist in the models of \cite{brinkman2023}, it is apparent that subdividing the parameter space by mass transfer case captures the dominant effect of altering the orbital period.

The apparent dominance of the evolutionary state of the donor star over the precise orbital period implies that it should be sufficient to compute a single exemplar case for each possible evolution state of the donor star. All that would remain would be to properly weight each mass transfer scenario in the binary population. Such weightings could be computed using binary population synthesis methods under a variety of physical assumptions to test the robustness of the weightings to different input physics. 

The most relevant mass transfer scenarios that would need to be computed to provide full coverage under this `exemplar case' framework would be:
\begin{itemize}
    \item main sequence mass transfer,
    \item post-main sequence mass transfer,
    \item mass transfer during core He burning,
    \item mass transfer after core He burning.
\end{itemize}

How altering the orbital period will affect yields from asymptotic giant branch stars remains unexplored, so it is uncertain whether a single (or small number of) exemplar case will be sufficient for each AGB progenitor mass. It is also possible that future studies similar to \cite{brinkman2023} that include supernova calculations and an expanded parameter space that includes episodes of mass transfer after core He burning will find that a full grid of orbital periods, rather than a smaller series of exemplar cases, is necessary to capture the full binary complexity. In this case, a methodology for including a range of orbital periods similar to what we describe in Section \ref{sec:effective_yields_better} could be used to include such a grid in the effective stellar yield calculation.

\subsubsection*{V. Complex binary evolution}

Having discussed the orbital period in detail, we now address the role of the initial mass ratio. The initial mass ratio will have a significant effect not only on the Roche geometry of the binary, but also on the evolution of this geometry during mass transfer. For this reason, different secondary masses will result in different mass transfer profiles, leading to somewhat different post-mass transfer outcomes. Perhaps more significantly, however, the mass ratio is critical to whether the mass transfer proceeds in a stable or dynamically unstable manner, and therefore to whether there is a need to invoke common envelope evolution.

Common envelopes occur under conditions of dynamically unstable mass transfer and have two possible outcomes: stellar mergers and short-period binaries. The outcome depends on the pre-common envelope binary configuration and highly uncertain binary stellar physics unique to the common envelope phase \citep{ivanova2013common}, but ultimately boils down to whether or not there is sufficient orbital energy to eject the common envelope before the binary merges \citep{iben1993common}. The stellar merger scenario we have already addressed, but the second scenario, where the binary survives, presents a significant increase in complexity.

In the case of the binary surviving the common envelope event, then you immediately have an extra mass loss term: the ejected envelope. This is not difficult to deal with: simply calculate the yields based on the entire envelope of the donor star being ejected (see e.g. \citealt{osborn2023}). The true increase in complexity lies in the subsequent evolution of the binary. The stripped star may go on to interact with its companion again as it evolves to conclude its life as a stripped supernova or a white dwarf. Further, the secondary, should it survive any subsequent interaction (including the possibility of a supernova kick), is extremely likely to interact with the remnant during its own evolution.

Accounting for this complexity is prohibitive in a purely detailed approach using a conventional grid of detailed binary models. Therefore, a hybrid approach combining wholistic, but synthetic, binary evolution with detailed binary evolution yields is needed. It should be possible to compute detailed models with yield calculations for different parts of the evolution which could then be stitched together in different sequences to estimate the total binary yield. The most important sequences, along with weights to balance their relative importance, can be identified using binary population synthesis calculations.

Grids of fully stripped stars have been computed in the past \citep{laplace2020}, and in principle there is nothing preventing consideration of what happens to the stripped star if it loses even more mass to a companion. If the amount of material that is accreted from the stripped star is taken from binary population synthesis also, then computing grids of stellar models where a star accretes stripped-star-composition material becomes far more tractable. Finally, we should note that many complex evolution channels will end in a (possibly exotic) stellar merger; further, complex series of binary evolution do not necessarily need to invoke common envelope physics.

The monumental task of being able to account for detailed structural changes in arbitrarily complex binary stellar evolution channels should, at this point, be clear. We would like to conclude by noting that difficult does not mean impossible. While the challenge is large in scale, it can be readily divided into many small, achievable tasks that, when combined, form an impressively complete whole capable of accounting for a far more realistic level of diversity in stellar evolution.

\section{Conclusions}
\label{sec:conc}

In this work, we describe a framework through which stellar yields that account for binary stellar evolution can be incorporated into galactic chemical evolution calculations. This is achieved by pre-computing the effective stellar yield for a mixed population defined by an arbitrary binary fraction and set of accretion efficiencies. Using the binary and single stellar yields of \cite{farmer2023}, we present effective stellar yields for a solar metallicity binary population considering stable mass accretion onto a main sequence companion star. As more binary stellar models exploring different evolution channels become available, this framework can be readily expanded to better approximate the full diversity and complexity of binary stellar evolution.

Within the scope of the binary stellar yields of \cite{farmer2023}, we describe the process of building a mixed binary stellar population. We consider different ways of treating the composition of the accreted material, as well as the effect of varying the binary mass fraction $h$ and the accretion efficiencies \sbs{\beta}{winds}, \sbs{\beta}{RLOF}, and \sbs{\beta}{SN} on the effective stellar yields. In the absence of detailed nuclear yields accounting for altered nuclear processing on the secondary as a result of composition changes to seed isotopes, we put forth our `CMCL4' model as a suitable first-order approximation for the treatment of accreted material. This model transfers any net production or depletion present in the accreted material due to the primary's nuclear processing to the yield of the secondary, thereby preserving the primary's nuclear processing rather than artificially erasing it, as assuming birth composition material would do.

This model also includes special cases for the radioisotopes the light elements Li, B, and Be. For the radioisotopes, we use a comprehensive radioactive decay network and an estimate for the secondary's remaining lifetime after mass transfer to calculate the amount of each isotope present at the end of the secondary's life. For the light elements, we artificially reset the primary's net enrichment or depletion present in the accreted material to zero, adjusting the appropriate daughter isotopes to preserve mass conservation. This has the effect of allowing the secondary's evolution to override the primary's in the accreted material, and prevents net-negative absolute secondary yields.

Considering a select few isotopes (C-12, N-14, O-16, Ne-22, Ni-56, and Fe-60), we show the effect of varying the binary parameters on the effective net yields. Varying $h$ while keeping the accretion efficiencies fixed to plausible values results in significant variation in the effective net yields for all considered isotopes. The effect of varying the accretion efficiency (under the CMCL4 model) on the effective net yields is generally much smaller than that of varying the binary fraction.

For the binary parameters, we also show the effect of accounting for a mixed binary population on the elemental abundance ratios [$X$/H]. Higher values of $h$ generally act to drive up the abundance ratios of these isotopes, with variations as high as 0.5 dex possible depending on the isotope and the primary mass. The binary fraction-dependent variation is highest for low-mass progenitors, reducing to less than 0.2~dex for the most massive stars considered. Varying the accretion efficiencies results in significantly lower variation in the [$X$/H] abundance ratios. Of the different mass transfer channels, variation in \sbs{\beta}{RLOF} causes the most significant variation, up to 0.5~dex for lower-mass primaries, and less than 0.2~dex for massive primaries. \sbs{\beta}{winds} and \sbs{\beta}{SN} both result in variation less than 0.2~dex even for the high-mass binaries where they are most significant.

The framework described in this work can be readily expanded to include additional binary stellar evolution models covering other binary evolution channels as more of the parameter space gets filled in. We identify low-mass stellar evolution, mergers, nuclear post-processed supernova yields, and low-metallicity evolution calculations as areas of the parameter space in most need of additional modelling work. With so much of the binary parameter space remaining unexplored in terms of detailed stellar yields, conclusive statements on the role of binary evolution of the chemical evolution of real stellar environments will remain elusive. However, with the recent binary modelling works of \cite{farmer2023}, \cite{brinkman2023}, and \cite{nguyen2024}, we are finally making progress in this area after decades of being limited to single stellar evolution models. It is imperative to maintain this momentum, expand the scope of these studies, and publicly report nuclear yields wherever possible in binary modelling works. This will allow them to be incorporated into frameworks like the one described in this paper and propagated into our models for real stellar populations.

\begin{acknowledgements}
The authors wish to acknowledge useful discussions with Hannah Brinkman, Eva Laplace, Ashley Ruiter, Ivo Seitenzahl, Maria Bergemann, and Annachiara Picco. The authors also wish to thank the referee, Hannah Brinkman, for their constructive and detailed report.
The research leading to these results has received financial support
from the Flemish Government under the long-term structural Methusalem funding program by means of the project SOUL: Stellar evolution in full glory, grant METH/24/012 at KU Leuven. TK acknowledges the support of the Indian Institute of Technology, Kanpur, in providing the resources for this research. TK is also thankful to the National Science Foundation under Grant No. OISE-1927130 (IReNA), which supported the participation in SDSS-V/IReNA Science Festival, 3-7 April 2023, Leuven, Belgium, where the idea of this work was conceived. 

\end{acknowledgements}

\section*{Data Availability}

Sets of pre-computed, machine-readable tables of effective stellar yields computed using the yields of \cite{farmer2023} under different binary assumptions are available online here: \href{https://zenodo.org/records/16089226}{10.5281/zenodo.16089226}. File structure is described in Appendix \ref{sec:online_sup_mat}. Additional binary configurations are available upon request from the authors, and the source code can be found here: \href{https://github.com/Alex162/BEEST---Binary-Evolution-and-Effective-Stellar-yields-Tookit}{https://github.com/Alex162/BEEST---Binary-Evolution-and-Effective-Stellar-yields-Tookit}.

\bibliographystyle{aa} 

\bibliography{bibfile.bib}

\appendix

\onecolumn
\FloatBarrier
\section{CMCL variation for $\beta_{\rm SN}=0.3$}

\begin{figure}[h!]
\centering
\begin{subfigure}{0.49\columnwidth}
\centering
\includegraphics[width=0.99\columnwidth]{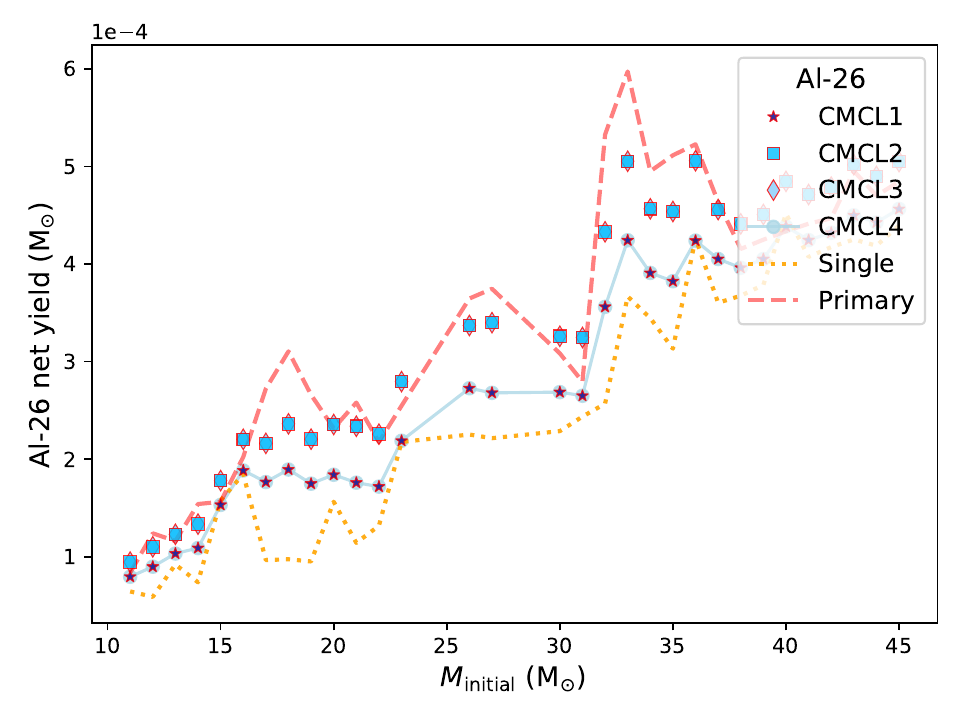}
\end{subfigure}%
\begin{subfigure}{0.49\columnwidth}
\centering
\includegraphics[width=0.99\columnwidth]{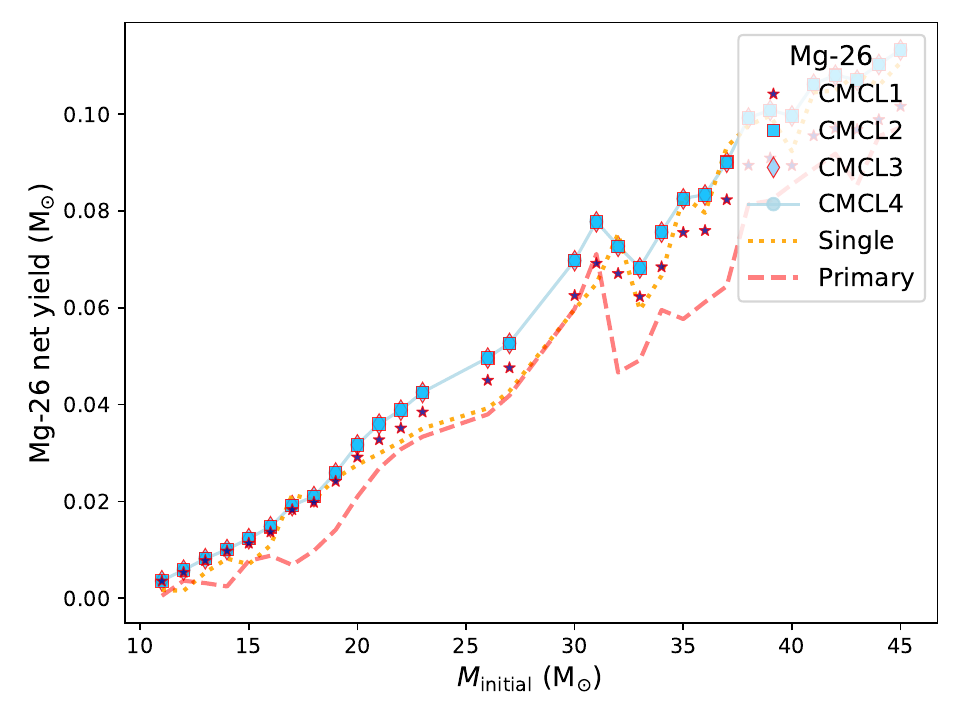}
\end{subfigure}

\begin{subfigure}{0.49\columnwidth}
\centering
\includegraphics[width=0.99\columnwidth]{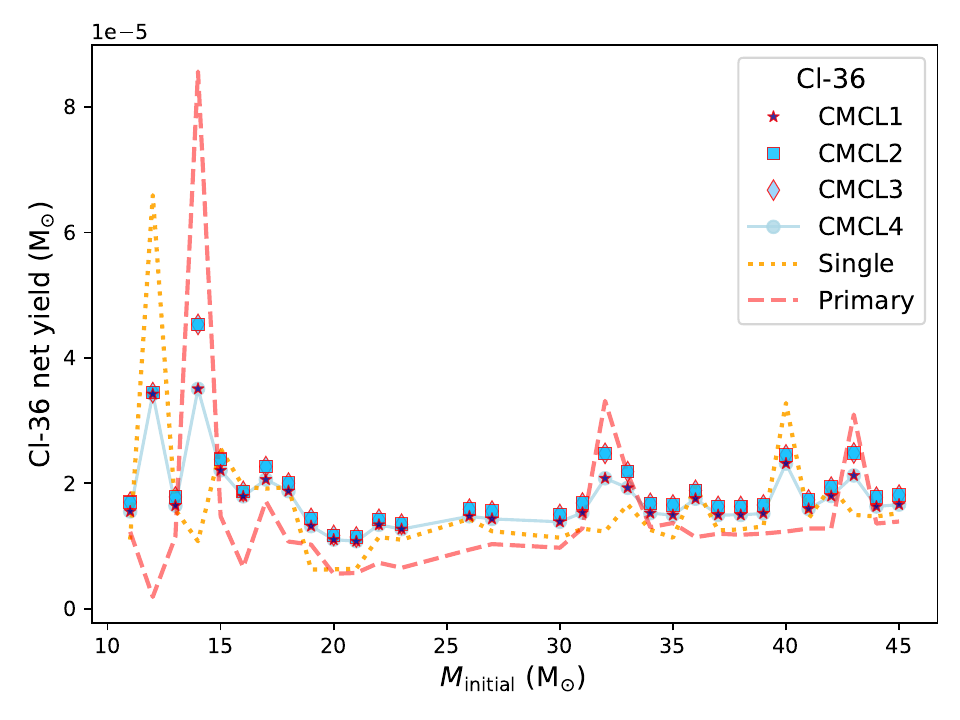}
\end{subfigure}%
\begin{subfigure}{0.49\columnwidth}
\centering
\includegraphics[width=0.99\columnwidth]{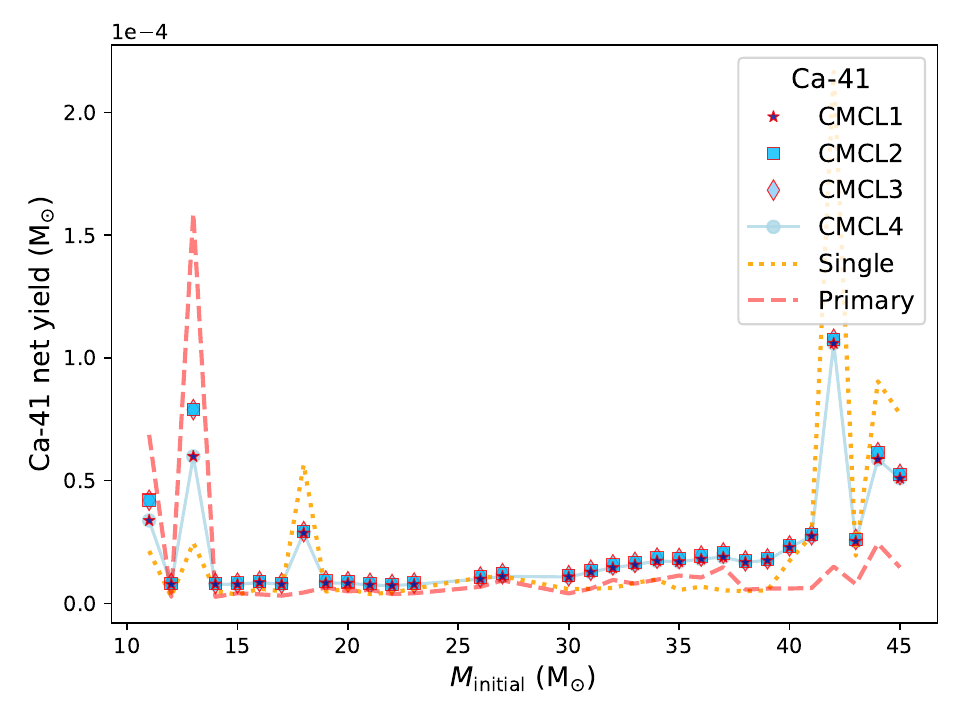}
\end{subfigure}

\begin{subfigure}{0.49\columnwidth}
\centering
\includegraphics[width=0.99\columnwidth]{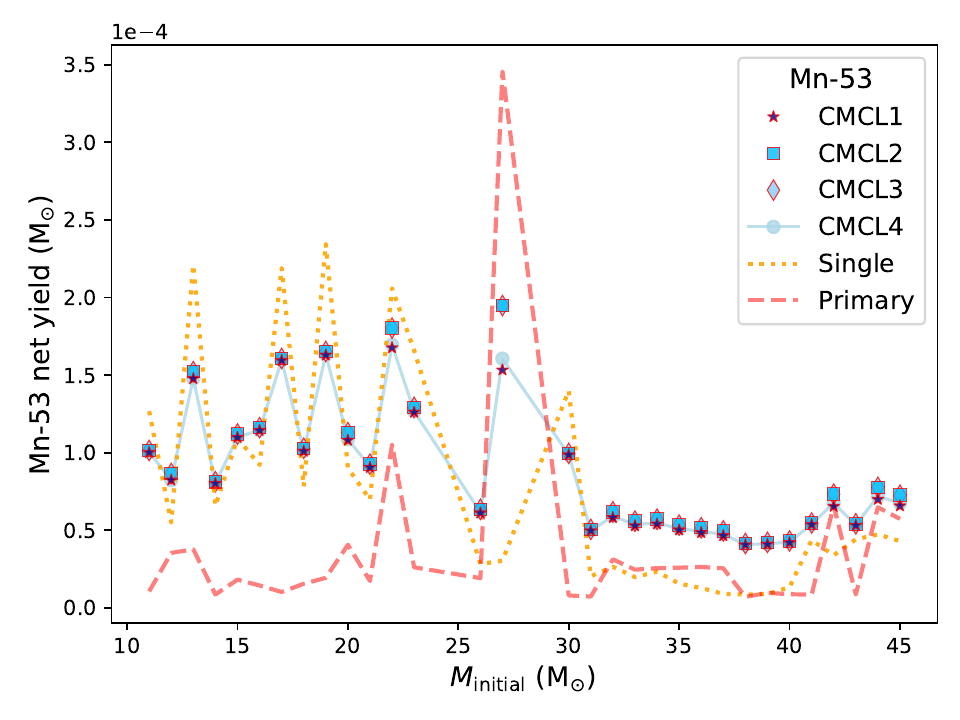}
\end{subfigure}%
\begin{subfigure}{0.49\columnwidth}
\centering
\includegraphics[width=0.99\columnwidth]{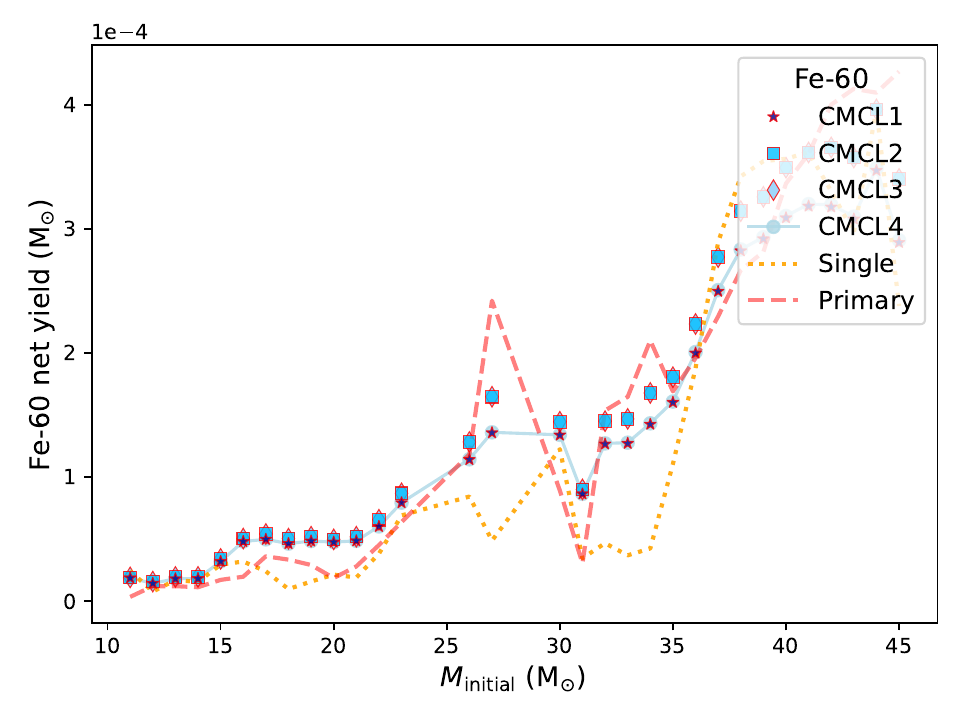}
\end{subfigure}

\caption{Net yields for the dominant isotopes of Al-26, Mg-26, Cl-36, Ca-41, Mn-53, and Fe-60. Effective yields computed for $h=0.5$, $\beta_{\rm winds}=0.1$, $\beta_{\rm RLOF}=1$, and $\beta_{\rm SN}=0.3$ are shown in shades of blue. The dashed lines present the net yields of each isotope for the reported single star (orange) and the reported primary star (red) for comparison.}
\label{fig:cmcl_radio_cc30}
\end{figure}

\pagebreak

\section{Online supplementary material}
\label{sec:online_sup_mat}
In the online supplementary material, we make available a number of pre-computed effective yield tables, as well as large numbers of additional figures, to assist GCE modelling efforts. The underlying python code will be made publicly available in the future, once it has been generalised for ease of use with a wider range of stellar models and proper levels of documentation have been implemented. In the mean time, if certain combinations of binary stellar parameters are required, they may be obtained on request from the authors.

In this section, we describe the file format and naming convention for the online supplementary material to ease navigation of this data set. The general file structure is:

\texttt{figs\textless A\textgreater\_\textless B\textgreater var\_\textless C\textgreater\_\textless D\textgreater\_\textless E\textgreater\_\textless F\textgreater\_radio/\textless filename\textgreater}

For example, the absolute effective yield table for C-13 computed with a binary fraction of 0.6, $\beta_{\rm winds}=0.1$, $\beta_{\rm SN}=0$, and $\beta_{\rm RLOF}=1$, the filename is \texttt{figsabs\_hvar\_winds0.1\_rlof1\_cc0\_radio/h\_0.6.txt} and the figure showing the variation in binary fraction for O-16 is:
\texttt{figsabs\_hvar\_winds0.1\_rlof1\_cc0\_radio/hvar\_o16.pdf}.

Further details on the filenames can be found in the sub-appendices, which deal with each component of the naming convention in detail.

The effective yield tables are plain-text files formatted for ease of use within the publicly available \texttt{OMEGA} GCE chemical evolution code \citep{cote2017omega,cote2018omegaplus}. This format is human-readable, and should be relatively easy to convert to input files for any GCE code capable of reading in yield tables from single stellar evolution models.

The global header contains information relating to the assumptions of the binary fraction and accretion efficiencies. Below this, each star is listed in sequence by its initial mass and metallicity, followed by the primary's lifetime in Myr and the (effective) remnant mass (M\solar). The first column less the second column is the net stellar yield; hence, for \texttt{net} tables, the second column is left blank. The remaining two columns give the proton and mass numbers for the isotope.

\subsection{\texttt{\textless A\textgreater}}

We provide two types of effective yield figures and tables: absolute \texttt{abs} and net \texttt{net}.

\subsection{\texttt{\textless B\textgreater}}

This indicates the binary or composition model complexity level (CMCL) parameter being varied. Possible values are: \texttt{h} (varying binary fraction $h$, \texttt{betawinds} (varying wind accretion efficiency $\beta_{\rm winds}$), \texttt{betarlof} (varying RLOF accretion efficiency $\beta_{\rm rlof}$), \texttt{betacc} (varying core-collapse supernova accretion efficiency $\beta_{\rm SN}$), and \texttt{cmcl} (varying composition model complexity). Note that for the CMCL files, we provide additional figures but not additional effective yield tables; for all provided effective yield tables, CMCL4 is adopted.
 
\subsection{\texttt{\textless C\textgreater,\textless D\textgreater,\textless E\textgreater,\textless F\textgreater}}

These positions describe the fixed values for parameters within this folder; as such, for all binary variation, only three of these positions will be used (the fourth variable is the one being varied). The order is always: $h$, $\beta_{\rm winds}$, $\beta_{\rm RLOF}$, and $\beta_{\rm SN}$ (\texttt{h}, \texttt{winds}, \texttt{rlof}, \texttt{cc}) with the fixed value immediately after the tag (e.g. \texttt{winds0.1}).

\subsection{\texttt{\textless filename\textgreater}}

The structure of the filename depends on whether it is a figure or a table. Tables are denoted by the variable parameter \texttt{\textless B\textgreater} followed by the value appropriate to that particular table and the .txt file extension (e.g. \texttt{h\_0.6.txt}). Figures are denoted by the variable parameter followed by \texttt{var\_\textless iso\textgreater.pdf}, where \texttt{\textless iso\textgreater} is the relevant isotope (e.g. \texttt{o16} for O-16). Possible values for the `variable' tag are the same as for \texttt{\textless B\textgreater}.

\begin{table*}[]
\caption{Exemplar table for \texttt{figsnet\_hvar\_winds0.1\_rlof1\_cc0\_radio/h\_0.5.txt}.}
\begin{tabular}{lllll}
\multicolumn{5}{l}{H   Effective binary yields for: h=0.5 beta\_winds=0.1, beta\_rlof=1, and beta\_cc=0} \\
\multicolumn{5}{l}{H Underlying yields: Farmer et   al., 2023} \\
\multicolumn{5}{l}{H Data prepared by: Alex Kemp,   Tejpreet Kaur,} \\
\multicolumn{5}{l}{H Isotopes: neut, H-1, H-2,   He-3, He-4, Li-7…, Zn-64} \\
\multicolumn{5}{l}{H Table:   (M=11.0,Z=0.0142)} \\
\multicolumn{2}{l}{H   Lifetime: 2.369e+01} &  &  &  \\
\multicolumn{2}{l}{H   Mfinal: 1.5252} &  &  &  \\
\&Isotopes & \&Yields & \&X0 & \&Z & \&A \\
\&H-1 & \&-4.235e+00 & \&0.000e+00 & \&1 & \&1 \\
\&H-2 & \&3.177e-13 & \&0.000e+00 & \&1 & \&2 \\
\&He-3 & \&-1.630e-04 & \&0.000e+00 & \&2 & \&3 \\
\&He-4 & \&7.898e-01 & \&0.000e+00 & \&2 & \&4 \\
… & … & ... & … & … \\
\&Li-7 & \&-9.466e-08 & \&0.000e+00 & \&3 & \&7 \\
\&Zn-62 & \&5.027e-04 & \&0.000e+00 & \&30 & \&62 \\
\&Zn-63 & \&4.366e-06 & \&0.000e+00 & \&30 & \&63 \\
\&Zn-64 & \&4.411e-05 & \&0.000e+00 & \&30 & \&64 \\
\multicolumn{3}{l}{H   Table: (M=12.0,Z=0.0142)} &  &  \\
\multicolumn{2}{l}{H   Lifetime: 2.044e+01} &  &  &  \\
\multicolumn{2}{l}{H   Mfinal: 1.6226} &  &  &  \\
\&Isotopes & \&Yields & \&X0 & \&Z & \&A \\
\&H-1 & \&-4.766e+00 & \&0.000e+00 & \&1 & \&1 \\
\&H-2 & \&5.892e-13 & \&0.000e+00 & \&1 & \&2 \\
\&He-3 & \&-1.871e-04 & \&0.000e+00 & \&2 & \&3 \\
\&He-4 & \&7.810e-01 & \&0.000e+00 & \&2 & \&4 \\
\&Li-7 & \&-1.033e-07 & \&0.000e+00 & \&3 & \&7 \\
… & … & … & … & … \\
\&Zn-62 & \&1.135e-03 & \&0.000e+00 & \&30 & \&62 \\
\&Zn-63 & \&1.545e-06 & \&0.000e+00 & \&30 & \&63 \\
\&Zn-64 & \&7.325e-05 & \&0.000e+00 & \&30 & \&64 \\
\multicolumn{3}{l}{H   Table: (M=13.0,Z=0.0142)} &  &  \\
\multicolumn{2}{l}{H   Lifetime: 1.80e+01} &  &  &  \\
\multicolumn{2}{l}{H   Mfinal: 1.5714} &  &  &  \\
\&Isotopes & \&Yields & \&X0 & \&Z & \&A \\
\&H-1 & \&-5.315e+00 & \&0.000e+00 & \&1 & \&1 \\
\&H-2 & \&4.658e-13 & \&0.000e+00 & \&1 & \&2 \\
\&He-3 & \&-2.109e-04 & \&0.000e+00 & \&2 & \&3 \\
\&He-4 & \&7.451e-01 & \&0.000e+00 & \&2 & \&4 \\
\&Li-7 & \&-1.119e-07 & \&0.000e+00 & \&3 & \&7 \\
... &... &... &... &... 
\end{tabular}
\tablefoot{The `Yields' column in this case lists the net yields, so the X0 column are 0. For the tables with absolute yields, X0 is the initial mass of each isotope and the net yield is given by subtracting X0 from the `Yields' column. The masses and yields are measured in solar masses, the lifetime is measured in Myr, and Z and A are the proton and mass numbers of each isotope, respectively.}
\label{tab:exemplar}
\end{table*}

\end{document}